\documentclass[showpacs,preprintnumbers,amsmath,amssymb]{revtex4}
\usepackage{graphicx}
\usepackage{bm}

\begin{document}

\title{
The classical dynamics of two-electron atoms near the triple collision
}
\author{ Min-Ho Lee$^{\dagger}$,
Gregor Tanner$^{\dagger\dagger}$ and
Nark Nyul Choi$^{\dagger}$
}

\affiliation{
$^{\dagger}$School of Natural Science, Kumoh National Institute
of Technology, Kumi, Kyungbook 730-701, Korea,\\
$^{\dagger\dagger}$School of Mathematical Sciences,
University of Nottingham, University Park, Nottingham NG7 2RD, UK
}

\begin{abstract}
The classical dynamics of two electrons in the Coulomb potential of
an attractive nucleus is chaotic in large parts of the high-dimensional
phase space. Quantum spectra of two-electron atoms, however, exhibit
structures which clearly hint at the existence of approximate symmetries in
this system. In a recent paper 
({\em Phys.\ Rev.\ Lett. {\bf 93}, 054302 (2004)}),
we presented a study of the dynamics near the triple collision as a first step
towards uncovering the hidden regularity in the classical dynamics of two
electron atoms. The non-regularisable triple collision singularity
is a main source of chaos in three body Coulomb problems. Here, we will give
a more detailed account of our findings based on a study of the global 
structure of the stable and unstable manifolds of the triple collision.
\end{abstract}

\maketitle

\section{Introduction}
\label{sec:intro}

Understanding the gravitational three-body problem as the simplest non-trivial
many-body system is of prime importance when considering dynamical
properties of the solar system such as its long term stability. Poincar\'e's
proof of the non-integrability of the three-body problem in 1890 showed that
closed form solutions of many-body systems are the exception rather than the
rule. This insight stood at the beginning
of modern dynamical systems theory concerned with developing tools to
understand the structures and stability properties of nonlinear dynamics.
Still, more than hundred years later, we know remarkably little about the
dynamics of three-body problems due to the large dimensionality of the
system, the long range interactions and the complexity of the dynamics near
the non-regularisable triple collision; see \cite{Dia97}, for a well written
account of the history of celestial dynamics before and after Poincar\'e's
discovery.

The microscopic counterpart of planetary motion, the dynamics of electrically
charged particles, occurs naturally in atoms and molecules; it has thus mainly
been studied in the context of quantum mechanics. First attempts to analyse the
classical dynamics of many-body Coulomb systems such as
two-electron atoms have been undertaken by the founding fathers of
quantum mechanics in order to extend Bohr's hydrogen quantisation rules
to more complex atoms. The failure to do so and the discovery of
Schr\"odinger's equation brought this project to an abrupt halt
in 1925. Only a better understanding of the use of semiclassical
methods for non-integrable systems pioneered by Gutzwiller and others
\cite{Gut90} in the 1970ies brought three-body Coulomb systems back onto the
agenda. These efforts led to the successful semiclassical description of
parts of the helium spectra in terms of collinear subspaces of the full
three-body problem in the 1990ies \cite{Ezr91,RW90a}.
Surprising regularities and selection rules in the spectrum of two
electron atoms, which have puzzled atomic physicist for decades, could now
be explained in terms of stability properties of an underlying
classical dynamics; see the review \cite{TRR00} for more
details.

Advances in a semiclassical treatment of the three-body Coulomb
problems were possible only due to a better understanding of the
classical dynamics in these systems. The existence of a perfect
Smale horseshoe giving rise to a complete binary symbolic dynamics
were uncovered for the collinear configuration where the two electrons
are on different sides of the nucleus (the $eZe$ configuration)
\cite{Ezr91,RTW93,TRR00,Yu98,Sano04}. Such a behaviour is a rare feature in
physically relevant dynamical systems and is here intricately linked to
the presence of the non-regularisable triple collision.
The collinear phase space where both electrons stay on the same side of the
nucleus - the $Zee$ configuration - has been found to be largely stable
in the full 5 dimensional phase space for $1 < Z < 10$
\cite{RW90a, TRR00, Duan00}.

Studies of the dynamics beyond the collinear configurations have so far
remained rare \cite{Gru95, YK98}. Quantum mechanical calculations \cite{Her83,
Bue95} suggest, however, that two-electron atoms have a variety of
approximate symmetries which express themselves in the form of approximate quantum
numbers in spectra of these atoms. This has been explained
qualitatively by group theoretical arguments \cite{Her83} and in terms of
adiabatic invariants \cite{Fea86}, see \cite{TRR00} for an overview. It is thus
only natural to ask how the existence of such approximate symmetries is reflected
in the classical dynamics of the corresponding three-body Coulomb problem.

Recently, we presented an analysis of the classical dynamics near the triple
collision in two-electron atoms in the full $L=0$ phase space \cite{CLT04}. The
triple collision and associate collision manifolds
are the key in understanding the structure of the dynamics of the five
dimensional phase space. Here, we will give a more detailed account of
the surprising effects observed in classical scattering signals below the three
particle breakup energy as well as how these effects arise due to the topology
of the phase space and the particle exchange symmetry.

The paper is organised as follows: in sec.\ \ref{sec:eq} we introduce the
McGehee scaling technique in hyperspherical coordinates. In sec.\ \ref{sec:E=0},
we describe the structure of the collision manifolds in the phase space for
$E=0$ which turns out to be relatively simple. We then treat the dynamics near
the triple collision for $E<0$ in sec.\ \ref{sec:E<0}  and we
present scaling laws similar to Wannier's threshold law \cite{Wa53} in some
detail. In the Appendix, we give the equations of motions 
combining Kustaanheimo-Stiefel transformation with McGehee scaling and discuss
the properties of a specific surface of section used in the main text.

\section{Equations of motion}
\label{sec:eq}

The classical three-body system can be reduced to four degrees of freedom
after eliminating the centre of mass motion and incorporating the
conservation of the total angular momentum.  We will focus here on
the special case of zero angular momentum, for which the motion of the
three particles is confined to a plane fixed in configuration space
\cite{Par65} and the problem reduces to three degrees of freedom, that is,
a five dimensional phase space for fixed energy. We will as usual work 
in the infinite nucleus mass approximation; the Hamiltonian including 
finite nuclear mass terms can be found in \cite{RTW93, Sano04}. In the 
following we will use scaling properties in
the three body Coulomb problem in two different ways: firstly, by scaling
the phase space coordinates with respect to energy and secondly, by
scaling out an overall size parameter thus considering only the shape
dynamics of the system.

By choosing a scaling transformation with respect to the total energy $E$
according to
\begin{equation} \label{scal_traf}
{\bf r}_{i} =  |E|\, {\bf r'}_{i}, \quad {\bf p}_i =
\frac{1}{\sqrt{|E|}} \; {\bf p'}_i
\end{equation}
where ${\bf r}_i, {\bf p}_i$ refer to the new coordinates and momenta of
electron $i = 1$ or 2, respectively, and introducing a time transformation
\begin{equation} \label{time_traf} t = \sqrt{|E|^3}\,  t' \, ,
\end{equation}
one deduces the new equations of motion from the Hamiltonian
\begin{equation} \label{He_scal}
H =  \frac{{\bf p}_1^2}{2}  + \frac{{\bf p}_2^2}{2}
- \frac{Z}{r_{1}} - \frac{Z}{r_{2}} + \frac{1}{r_{12}} =
\left\{
\begin{array}{rcl}
+1 & : & E > 0 \\
0 & : & E = 0 \\
-1 & : & E < 0
\end{array} \right. \, .
\end{equation}
Here, $Z$ refers to the charge of the nucleus (in units of the elementary
charge) and masses are given in units of the mass of the electron.
We will in general consider $Z=2$, that is, Helium, if not specified
otherwise. Furthermore, $r_i, r_{12}$ denotes the nucleus-electron and
electron-electron distances, respectively.

From eqn.\ (\ref{He_scal}) it is clear that we only have to consider three
different values of the energy. Our ultimate goal is to better understand the
bound and resonance states in quantum two-electron atoms and we are
thus most interested in the classical dynamics for $E < 0$, that is,
we will consider $E = -1$. In this regime, only one electron can escape
classically and it will do so for most initial conditions. It turns out,
however, that one can learn a lot about the $E < 0$ - dynamics by
analysing the dynamics at the three-particle breakup threshold
$E = 0$ in detail. The phase space can be reduced to 4 dimensions
in this case and the dynamics in the reduced space turns
out to be relatively simple as will be discussed in section
\ref{sec:E=0}. A similar approach has been employed by Wannier
\cite{Wa53}; by extrapolating dynamical behaviour at $E=0$ to
the dynamics for $E >0$, he was able to deduce his celebrated threshold
law for the total two-electron ionisation cross section which turns out to
be completely classical in nature \cite{Ros98}.
 
How are the spaces $E = \pm 1$ and $E=0$ connected? When considering
scattering trajectories where one electron, say electron 1, approaches the
nucleus from $r_1 = \infty$ with energy $E_1$, the energy scaling property,
eqs.~(\ref{scal_traf}), (\ref{time_traf}),  implies that the dynamics depends
on the ratio $E / E_1$ only rather than on the absolute values
of $E$ and $E_1$ separately. The limit $E\to 0$ is thus equivalent to
$E / E_1 \to 0$ which can for fixed $E = E_1 + E_2 = \pm 1$
be achieved by for example considering the limit $E_1 \to \infty$. (In the
same way, we may consider the limit $E\to 0$ for fixed $E_1$).
As we will see in sec.~\ref{sec:E<0}, the limit $E \to 0$ is also closely
related to the dynamics near the triple collision.

The dynamics for $E = 0$ can be reduced to 4 dimension using an
additional scaling relation. Following
McGehee \cite{McG74}, one uses the similarity of the
overall dynamics when rescaling the total size of the system. This
means that the shape dynamics given by the relative positions of the three
particles in space decouples from the overall change in size
of the system in certain limits.
We introduce the hyperradius $R = \sqrt{r_1^2 + r_2^2}$ as
an overall scaling parameter and shape parameters given by
the hyperangle
$\alpha = \tan^{-1} \left({r_2}/{r_1}\right)$
and the inter-electronic angle
$\theta  = \angle({\bf r}_1,{\bf r}_2) = \theta_1 - \theta_2$
with $\theta_i$ being the azimuthal angles. The Hamiltonian
(\ref{He_scal}) written in these hyperspherical coordinates
has for angular momentum $L =0$ the form
\begin{eqnarray}\label{He_shape}
H &=&
\frac{1}{2} \left( p_{r_1}^2 + \frac{p_{\theta_1}^2}{r_1^2} +
                   p_{r_2}^2 + \frac{p_{\theta_2}^2}{r_2^2}
            \right)
+ \frac{1}{R}V(\alpha,\theta) \nonumber \\
&=&
\frac{1}{2} \left( p_R^2 + \frac{p^2_{\alpha}}{R^2}
+ \frac{p^2_{\theta}}{R^2\cos^2\alpha \sin^2\alpha}\right)
+ \frac{1}{R}V(\alpha,\theta)
\end{eqnarray}
with
\[
V(\alpha,\theta) = -\frac{Z}{\cos\alpha} - \frac{Z}{\sin\alpha}
+ \frac{1}{\sqrt{1 - 2 \cos\alpha \sin\alpha \cos\theta}}\, .
\]
Note, that for $L=0$, we have
\begin{equation}\label{angular}
p_\theta = p_{\theta_1} = - p_{\theta_2},
\end{equation}
where $p_{\theta}$ is the momentum conjugate to the inter-electronic angle $\theta$.
The triple collision corresponds to $R = 0$, here.
For Hamiltonians of the form (\ref{He_shape}), one can separate the
shape dynamics from the overall scale dynamics given by the time dependence of
the hyperradius $R(t)$. Such a separation is exact for $E=0$ and
reflects the dynamics in the limit $R \to 0$, that is, close
to the triple collision for $E \ne 0$. In analogy with (\ref{scal_traf}),
(\ref{time_traf}), one defines the (time-dependent) scaling transformation
\begin{eqnarray} \label{scal_R}
\bar{\alpha} = \alpha; \qquad&\quad \bar{\theta} = \theta;&
                \quad \bar{R}  = \frac{1}{R}\, R = 1;\\ \nonumber
\bar{p}_R=\sqrt{R}\, p_R; &\quad \bar{p}_{\alpha}=\frac{1}{\sqrt{R}}\, p_{\alpha};&
\quad \bar{p}_{\theta} = \frac{1}{\sqrt{R}}\, p_{\theta};\\ \nonumber
d\bar{t} = \frac{1}{R^{3/2}}\, dt;& \quad \bar{H} = \bar{E} = R\, E\, .&
\end{eqnarray}
\noindent
Note that the above transformations are invariant under rescaling the energy
according to (\ref{scal_traf}), (\ref{time_traf}), that is, it is again
sufficient to consider the case $E=\pm 1$ or $0$ only.
The transformations (\ref{scal_R}) do, however, destroy the symplectic
structure of the original differential equations; the new Hamiltonian $\bar{H}$
is in particular no longer a constant of motion for $E\ne 0$. The equations
of motion with respect to the rescaled time are
\begin{eqnarray} \label{EoM}
\dot{\alpha} = p_{\alpha}; \qquad \quad \;\;\;\;
&\quad \dot{p}_{\alpha}& = -\frac{1}{2}\, p_R\, p_{\alpha} +
p_{\theta}^2\;
\frac{\cos^2\alpha - \sin^2\alpha}{\sin^3\alpha \cos^3\alpha} -
\frac{\partial}{\partial \alpha} V(\alpha,\theta);\\ \nonumber
\dot{\theta} = \frac{p_{\theta}}{\sin^2\alpha \cos^2\alpha};
&\quad \dot{p}_{\theta}& =
- \frac{1}{2}\, p_R\, p_{\theta} -
\frac{\partial}{\partial \theta} V(\alpha,\theta);\\\nonumber
\dot{\bar{H}} = p_R  \bar{H}; \qquad \;\;\;\;
& \quad \dot{p}_R &= \frac{1}{2}\, p_{\alpha}^2 +
\frac{1}{2} \frac{p_{\theta}^2}{\cos^2\alpha \sin^2\alpha} +
{\bar{H}}
\end{eqnarray}
with
\begin{equation}\label{H_McGehee}
\bar{H} = R H = \frac{1}{2} \left( p_R^2 + p_{\alpha}^2 +
\frac{p_{\theta}^2}{\cos^2\alpha \sin^2\alpha} \right) +
V(\alpha,\theta)= R E \,
\end{equation}
where we skip the bar signs again for convenience except for $\bar{H}$.

The new equations of motion (\ref{EoM}) are indeed independent of $R$;
the explicit time dependence of $R(t)$ can be recovered from (\ref{H_McGehee})
for $E\ne 0$ or may be obtained by integrating
$\dot{R} = p_R R$ along a trajectory for $E=0$.

The problem simplifies when considering the special initial
condition $\bar{H} = 0$, that is, $E = 0$ \underline{or} $R = 0$.
Firstly, a true reduction in dimensionality is achieved as $\bar{H}$
becomes a constant of motion and we are left with only four independent
coordinates. Secondly, for $\bar{H} = 0$ we have $\dot{p}_R \ge 0$, and
the scaled momentum $p_R$ increases monotonically with time. This leads to
a relatively simple overall dynamics in the $\bar{H} = 0$ subspace
which will be studied in detail in the next section.

The triple collision itself has been lifted from the equations of motions
(\ref{EoM}) by the time transformation in (\ref{scal_R}). Two fixed points
are created instead which are related to the triple collision.
These fixed points can not be reached in finite (scaled) time which is
a manifestation of the non-regularisability of the triple collision singularity.
Other singularities are still present in (\ref{EoM}), the binary collisions
at $r_i = 0$ or equivalently at $\alpha = 0$ or $\pi/2$. They can be
regularised by standard techniques such as Kustaanheimo-Stiefel (KS)
transformation \cite{KS65,AZ74,RTW93}. A set of singularity-free
equations of motions is obtained by first employing a KS - transformation
using parabolic coordinates and then using McGehee's scaling technique for
this new set of coordinates. Details of the derivation can be found in
Appendix \ref{asec:KS}; the resulting differential equations
(\ref{appendix:EoP}), (\ref{appendix:scaledEoQ}), and (\ref{appendix:scaledE})
have been used throughout the paper for numerical calculations. The
description in terms of parabolic coordinates is, however, less transparent
than the hyperspherical coordinates and the latter are thus used
in the discussion of the dynamics.

At a binary collisions $\alpha = 0$ or $\alpha = \pi / 2$, the value of
$p_\alpha$ makes an instantaneous transitions from $\mp \infty$ to
$\pm \infty$ whereas all other variables behave smoothly
at these points. We may thus identify $p_\alpha$ before and after the
collision by introducing the regularised variable
\begin{equation} \label{reg_pa}
\bar{p}_{\alpha} = p_\alpha \sin 2\alpha.
\end{equation}
The resulting set of smooth hyperspherical coordinates including the
regularised $\bar{p}_{\alpha}$ will be used in our description of the
phase space structures. Note, however, that in contrast to the case
of the collinear $eZe$ subspace \cite{McG74,Yu98,Sano04},
(\ref{reg_pa}) can not be used to remove the binary collision singularities
in the equations of motions. Instead, one has to go to parabolic
coordinates to obtain a set of fully regularised ODE's as presented in
Appendix~\ref{asec:KS}.

\section{Dynamics for $E = 0$} \label{sec:E=0}
\subsection{Fixed points and invariant subspaces} \label{sec:Invariant}
In order to understand the dynamics near the triple collision for
$E < 0$ it is advantageous to analyse the topology of the flow generated
by (\ref{EoM}) for $E = 0$.
We start by briefly discussing the fixed points and the invariant
subspaces of the dynamics in the rescaled coordinates. For $E=0$,
the dynamics takes place on a 4 dimensional manifold in a 5
dimensional space. There are two fixed points of the flow, that is,
\[\alpha = \pi/4, \quad \theta = \pi,
\quad p_{\alpha}=0, \quad p_{\theta} = 0,
\quad p_R = \pm \sqrt{\sqrt{2}(4 Z -1)} = \pm P_0 .\]
These fixed points correspond to trajectories in the
full phase space where both electrons fall into the nucleus
symmetrically along the collinear axis, that is, the
{\em triple collision point} (TCP) with $p_R = - P_0$ and
its time reversed partner, the trajectory of symmetric double
escape, that is, the {\em double escape point} (DEP) with
$p_R = P_0$. In addition, there are three invariant subspaces:
the collinear spaces $ \theta = \pi$, $p_{\theta} = 0$
(the $eZe$ configuration) and  $\theta = 0$, $p_{\theta} = 0$
(the $Zee$ configuration) and the so-called
Wannier ridge (WR) of symmetric electron dynamics with
$\alpha = \pi/4$, $p_{\alpha} = 0$.

\begin{figure}
\includegraphics[scale=0.50]{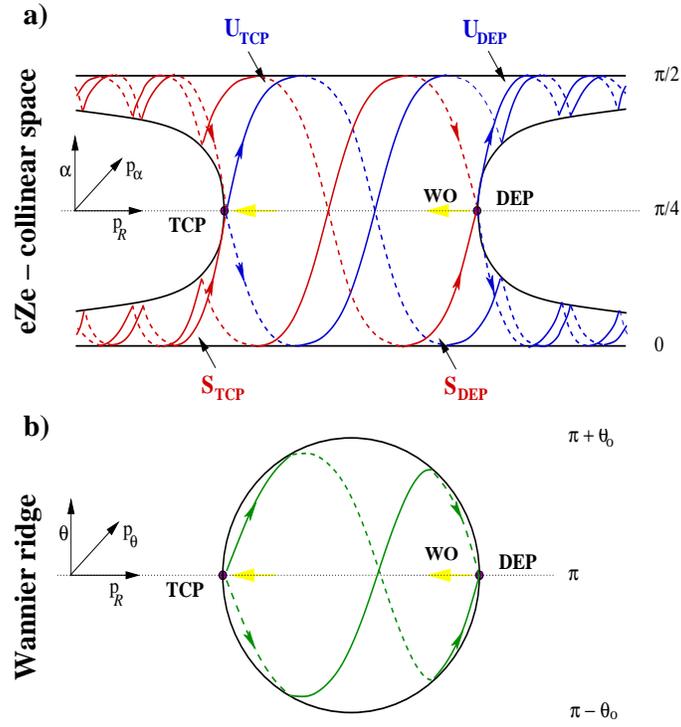}
\caption[]{\small The $eZe$ manifold (a) and the Wannier ridge manifold (b)
for $E=0$. The angle $\theta_0$ in (b) corresponds to the maximal deviation
from the collinear configuration $\theta = \pi$ possible in the WR for $E=0$
(in fact $\pi - \theta_0 = \arccos(1 - 1/8 Z^2)$.) The 2-dim.\ invariant
subspaces are embedded in the full phase space $E\le 0$ of dimension 5;
the subspaces are connected at the TCP and DEP (for $E=0$) and along the
Wannier orbit (WO) for $E<0$.
}
\label{fig:fig1}
\end{figure}

In the $eZe$ space with Hamiltonian
\begin{equation} \label{eZe-configuration}
\bar{H} = \frac{1}{2} (p_R^2 + p_\alpha^2 )
- \frac{Z}{\cos \alpha} - \frac{Z}{\sin \alpha}
+ \frac{1}{\cos \alpha + \sin \alpha} = 0 \, ,
\end{equation}
a typical trajectory represents an outer electron
coming from infinity with $p_R = -\infty$,
$\alpha = 0$ or $\pi/2$ and one of the two electrons leaving
towards infinity with $p_R \to \infty$, $\alpha \to 0$ or $\pi/2$.
Identifying the points $p_{\alpha} = \pm \infty$
at the binary collisions $\alpha = 0, \pi/2$ by using $\bar{p}_{\alpha}$ as
discussed in sec.\ \ref{sec:eq}, the topology of the $eZe$ - phase space
takes on the form of a sphere with four points taken to infinity, see
Fig.~\ref{fig:fig1}a \cite{Yu98, Sano04, McG74}. The two fixed points are 
located at the saddles between the arms stretching in forward and backward 
direction along the $p_R$ - axis. The $eZe$-space for $E<0$ fills the interior
of the manifold in Fig.~\ref{fig:fig1}a

The Wannier ridge space described by
\begin{equation} \label{WR-configuration}
\bar{H} = \frac{1}{2} p_R^2 + 2 p_\theta^2 - 2 \sqrt{2} Z
+ \frac{1}{\sqrt{1 - \cos \theta}} = 0 \, ,
\end{equation}
is, on the other hand, a compact space with the topology of a sphere
where the fixed points form opposite poles, see Fig.~\ref{fig:fig1}b. 
The dynamics for $E=0$ is trivial as the full space acts as the unstable 
manifold of the TCP as well as the stable manifold of the DEP.
The interior of the sphere corresponds to the phase space of the 
WR   for $E < 0$. The dynamics is of mixed type containing stable islands
and ergodic regions for $Z > 1/4$. In what follows we will not discuss the 
features of the WR-dynamics in more detail, see \cite{RW90, TRR00} for 
details as well as \cite{Dia04} for a more rigorous approach.
Note, that the $eZe$ configuration and the  WR  are connected at the
fixed points (in $E = 0$) and along the so called Wannier orbit (WO)
or symmetric stretch orbit with $\alpha=\pi/4$, $\theta = \pi$, 
$p_{\alpha} = 0$ and $p_{\theta} = 0$ with $E < 0$.

The overall dynamics is invariant under
the transformation $p_j \to -p_j$ and $dt \to -dt$ with $j = R, \theta$
or $\alpha$ reflecting the time-reversal symmetry of the original problem.
The triple collision point and double escape point are thus equivalent and
related by time reversal symmetry.

The linearised dynamics near the fixed points can be obtained
directly from the equations (\ref{EoM}); for each fixed point, two of
the four eigenvectors in $E=0$ lie in the $eZe$ space, the other
two on the Wannier ridge. One obtains in particular for the eigenvalues
at the TCP \cite{Wa53}
\begin{eqnarray} \label{stab_eZe}
&\lambda_{S_{T}}^{eZe}& = \frac{P_0}{4}
\left(1 - \sqrt{\frac{100 Z -9}{4 Z - 1}} \right) \quad
\;\mbox{eZe: stable}\\ \nonumber
&\lambda_{U_{T}}^{eZe}& = \frac{P_0}{4}
\left(1 + \sqrt{\frac{100 Z -9}{4 Z - 1}} \right) \quad
\;\mbox{eZe: unstable}\\ \nonumber
&\lambda_{U_{T}}^{WR}& = \frac{P_0}{4}
\left(1 \pm \sqrt{\frac{4 Z -9}{4 Z -1}} \right) \qquad
\;\mbox{Wannier ridge: unstable},
\end{eqnarray}
and for the DEP
\begin{eqnarray} \label{stab_DEP}
&\lambda_{U_{D}}^{eZe}& = -\frac{P_0}{4}
\left(1 - \sqrt{\frac{100 Z - 9}{4 Z -1}} \right) \quad
\mbox{eZe: unstable}\\ \nonumber
&\lambda_{S_{D}}^{eZe}& = -\frac{P_0}{4}
\left(1 + \sqrt{\frac{100 Z - 9}{4 Z -1}} \right) \quad
\mbox{eZe: stable}\\ \nonumber
&\lambda_{S_{D}}^{WR}& =
-\frac{P_0}{4} \left(1 \pm \sqrt{\frac{4 Z - 9}{4 Z -1}}\right) \qquad
\mbox{Wannier ridge: stable} .
\end{eqnarray}
The eigendirections leading out of the $\bar{H} = 0$ subspace
are directed along the $p_R$ axis; the corresponding stable and
unstable manifolds $S_{T}^{\bar{H}\ne 0}$, $U_{D}^{\bar{H}\ne 0}$
are embedded both in the $eZe$ and  WR  space and are thus identical
to the Wannier orbit.  That is, the WO forms a heteroclinic
connection leading from the DEP to the TCP.  The stabilities along
the eigendirections are
\begin{equation} \label{stab_H}
\lambda_{S_{T}}^{\bar{H}\ne 0} = -P_0; \qquad
\lambda_{U_{D}}^{\bar{H}\ne 0} = P_0
\end{equation}
and $P_0 = \sqrt{\sqrt{2}(4 Z -1)}$ as defined above.

Tab.~\ref{table:table1} gives an overview over how various parts of
the stable and unstable manifold of the fixed points are embedded
within the invariant subspaces. The TCP has, in particular, three
unstable directions and two stable directions of which one is coming
from outside the $E=0$ subspace, see also Fig.~\ref{fig:fig1}. The
converse holds for the DEP which has three stable directions all in
$E=0$ and two unstable directions.

The TCP can only be reached by trajectories on the 2-dim.\ stable
manifold of the TCP which is fully embedded in the $eZe$ space, see
Fig.~\ref{fig:fig1}a. Trajectories in $E = 0$ approaching the TCP
in the $eZe$ space close to the $S_{T}^{eZe}$ will leave the
neighbourhood of the TCP along the unstable manifold $U_{T}^{eZe}$
which leads to single ionisation of one of the electrons eventually.
The dynamics near the TCP is thus for $E = 0$ well
separated from the DEP and the two fixed points are dynamically not
connected. (Strictly speaking, this is true only for
$Z>0.287742...$; at the critical value the system is degenerate, that is,
$U_{T}^{eZe}$ coincides with $S_{D}^{eZe}$ \cite{Yu98}; this parameter regime
is, however, physically not relevant.)

The situation changes when leaving the $eZe$ space into the full
4 dimensional space $E=0$. The Wannier ridge itself provides
now a connection between the TCP and DEP and trajectories approaching
the TCP can leave along the Wannier ridge and thus come close to the DEP.
The 3 dimensional stable manifold $S_{D}$ of the DEP which contains
the Wannier ridge and the $S_{D}^{eZe}$ acts in fact as the
stable manifold of the Wannier ridge itself or more precisely
$S_{WR} = S_{D} \cup S_{T}^{eZe}$ and $S_{D}$ is
thus connected to $S_{T}^{eZe}$.
In what follows, the $S_{D}$ will be of special importance for
understanding some of the striking features in the classical
electron - impact scattering signal found for $E=0$, see
sec.\ \ref{sec:E=0signal}, as well as in the $E<0$ regime discussed
in detail in sec.\ \ref{sec:E<0}.

A summary of the submanifolds of the stable and unstable manifolds
of the fixed points and the spaces, they are embedded in, can be found
in table \ref{table:table1}. Note in particular that $U_T$ and $U_D$
are related to $S_D$ and $S_T$ by time reversal symmetry; thus,
$U_T$ together with $U_{D}^{eZe}$ form the unstable manifold
of the Wannier ridge, $U_{WR}$, in $E=0$.

\begin{table}
\begin{tabular}{|l|c|c|c|c|c|c|c|c||c|c|c|c|}  \hline
      & $S^{eZe}_{T} $
      & $S^{\bar{H} \neq 0}_{T} $
      & $U^{eZe}_{T} $
      & $U^{WR}_{T} $
      & $S^{eZe}_{D} $
      & $S^{WR}_{D} $
      & $U^{eZe}_{D} $
      & $U^{\bar{H} \neq  0}_{D} $
      & $S_T$
      & $U_T$
      & $S_D$
      & $U_D$
      \\ \hline
Dimension
      & 1
      & 1
      & 1
      & 2

      & 1
      & 2
      & 1
      & 1
      & 2
      & 3
      & 3
      & 2
      \\ \hline
Embedded in
      & $eZe $
      & $eZe $
      & $eZe $
      &
      & $eZe $
      &
      & $eZe $
      & $eZe $
      & $eZe $
      &
      &
      & $eZe $
      \\
      &
      & $WR$
      &
      & $WR$
      &
      & $WR$
      &
      & $WR$
      &
      &
      &
      &
      \\
      & $\bar{H} = 0 $
      & $\bar{H} \ne 0 $
      & $\bar{H} = 0 $
      & $\bar{H} = 0$
      & $\bar{H} = 0$
      & $\bar{H} = 0 $
      & $\bar{H} = 0$
      & $\bar{H} \ne 0$
      &
      & $\bar{H} = 0$
      & $\bar{H} = 0$
      &
      \\ \hline
\end{tabular}

\caption[]{Dimensions and embedding spaces of invariant subspaces of the
stable/unstable manifolds of the fixed points TCP and DEP .
}
\label{table:table1}
\end{table}

\subsection{The stable manifold of the DEP} \label{sec:SDEP}
We analyse first the topology of the 4-dimensional invariant
subspace $\bar{H}=0$ which is most conveniently studied by
considering the 3-dimensional Poincar\'e surface of section (PSOS)
$\theta = \pi$, $\dot{\theta} \ge 0$ in $\alpha$ -
$\bar{p}_{\alpha}$ - $p_R$ coordinates.
The surface $\theta = \pi$ is indeed a good PSOS in the sense
that the flow is not tangential to the surface except
for trajectories in the $eZe$ space which is an invariant subspace
fully embedded in the PSOS; the $eZe$ forms in fact
the boundary of the surface of section as can be seen from eqs.\
(\ref{H_McGehee}) and (\ref{eZe-configuration}). In addition,
almost all trajectories cross the surface at least once, see
Appendix~\ref{appendix:sec:PSOS} for details.

\begin{figure}
\includegraphics[scale=0.48]{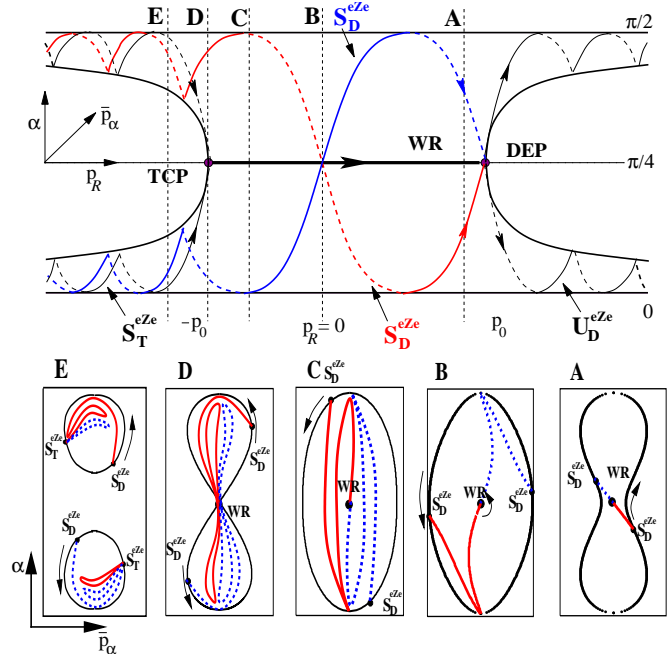}
\caption[]{\small  The PSOS $\theta = \pi$ in the
$E=0$ subspace in $\alpha$-$\bar{p}_{\alpha}$-$p_R$
coordinates. The $eZe$ space forms the boundary
of the PSOS, the WR connects the TCP and DEP
along the $p_R$ axis at $\alpha = \pi/4$, $p_{\alpha} =0$.
Various cuts of the PSOS at fixed $p_R$ - values
together with the $S_D$ are shown below. The two arms of the
$S_D$ stretching from the WR towards the $S^{eZe}_D$ on
the $eZe$ boundary are shown as full and dashed line, respectively.
(The cuts {\bf C} -- {\bf E} are drawn schematically to
enhance important features.)
}
\label{fig:sdep}
\end{figure}

The PSOS has in $\alpha$ - $\bar{p}_{\alpha}$ - $p_R$
coordinates the form of the $eZe$ space in Fig.\ \ref{fig:fig1}a.
The interior of the 2-dim.\ $eZe$ manifold represents here, however,
the domain of the Poincar\'e map $\theta = \pi$ for $p_{\theta} \ge 0$
and $E = 0$, see Fig.\ \ref{fig:sdep}. 
The fixed points TCP and DEP lie on the boundary
of the PSOS, whereas the 2-dimensional Wannier ridge space
in the PSOS forms a line connecting the TCP and DEP
along the $p_R$ - axis at $\alpha = \pi / 4$, $p_{\alpha} = 0$.

Due to $\dot{p}_R \ge 0$ in (\ref{EoM}), $p_R$ increases
monotonically with time leading to a relatively simple overall dynamics in
$\bar{H}=0$. Its important features can be characterised by the behaviour
of the stable/unstable manifolds of the fixed points. Especially, the
co-dimension one manifold $S_D$ is a good candidate for supplying a
dividing surface in the full $E=0$ phase space. In Fig.~\ref{fig:sdep},
the topology of the $S_D$ in the PSOS is discussed by showing
cuts through the PSOS at fixed $p_R$ values with $p_R <P_0$.

The $S_{D}$ is for $-P_0 \le p_R \le P_0$ bounded by the
1-dimensional stable manifold $S_{D}^{eZe}$ in the $eZe$ -
space, and the 2-dimensional Wannier ridge. Remarkable is the
evolution of this manifold near the TCP at $p_R = - P_0$, 
where the phase space itself splits
into two distinct parts. Starting at the DEP fixed point
at $p_R = P_0$, we will discuss the form of $S_D$ by
going towards decreasing $p_R$ - values which corresponds essentially
to an evolution of the $S_{D}$ backward in time.
The $S_{D}$ undergoes the usual stretching and folding mechanism
typical for an unstable manifold in bounded domains.
The stretching and folding is here facilitated by an overall
rotation of the space around the Wannier ridge - axis
$\alpha = \pi / 4$, $p_{\alpha} = 0$ and a certain "stickiness"
near $\alpha = 0$ or $\pi / 2$
(see the cuts {\bf B} and {\bf C} in Fig.~\ref{fig:sdep}).
The behaviour near the binary collision points is due to our
choice of regularised momentum $\bar{p}_{\alpha}$ which projects
the phase space at $\alpha = 0$ or $\pi / 2$ onto the point
$\bar{p}_{\alpha} = 0$.

As $p_R$ moves towards the TCP at $-P_0$, the phase space develops a
bottle neck whereas the $S_{D}$ stretches over the whole phase space 5
times by now. That means, that as $p_R$ decreases further passing
through $- P_0$, the $S_{D}$ is cut at the TCP into distinct parts,
see {\bf D} in Fig.~\ref{fig:sdep}. We end up with 5 pieces of the
$S_D$ in each arm. The only way to leave the TCP (backward in time)
is along the stable manifold $S_{T}^{eZe}$ in the $eZe$ space. This
implies that the 5 pieces in each arm are connected at the
$S_{T}^{eZe}$ for $p_R < -P_0$ forming two loops and one connection to the
$eZe$ boundary at $S_D^{eZe}$, see Fig.~\ref{fig:sdep} ({\bf E}).  The
$S_{T}^{eZe}$ itself is thus a boundary of the $S_{D}$ without being a
part of it and $S_{D}$ connects the stable manifolds $S_{T}^{eZe}$
and $S_{D}^{eZe}$ for $p_R < -P_0$.

\begin{figure}
\includegraphics[scale=0.58]{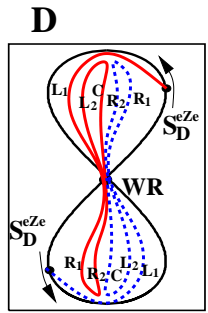}
\caption[]{\small The 5 pieces of the $S_{D}$ at the cut ${\bf D}$
in Fig.~\ref{fig:sdep} are label $R_1$, $R_2$, $C$, $L_2$ and $L_1$
as indicated in the figure.
}
\label{fig:numbering}
\end{figure}

There are two main routes to approach the DEP for electrons
coming in from $p_R = -\infty$ close to the $eZe$ - boundary:
firstly, a trajectories can approach the DEP 'directly' by moving
in the vicinity of the $S_{D}^{eZe}$; this is the only path open in
the eZe space. In the full $E=0$ space, a second route opens up;
trajectories close to the $S_{T}^{eZe}$ approaching the TCP
can stay close to one of the 5 leaves of the $S_{D}$ and move
along the $S_D$ towards the
DEP. This twofold approach turns out to be the main new element when moving
away from the collinear spaces. For later reference, we will label the leaves
of the $S_D$ according to $R_1$, $R_2$, $C$, $L_2$ and $L_1$
as indicated in Fig.~\ref{fig:numbering}. Note that the central leaf (C)
is the one connected directly to the WR for $p_R > -P_0$, whereas the
leaves to the right, $R_{1,2}$, and to the left, $L_{1,2}$, do not stay close
to the WR when leaving the TCP.

Fig.~\ref{fig:sdep} is based on numerical calculations
for $Z=2$; no changes in the topological structure and in the number
of leaves of the $S_{D}$ are recorded for nuclear charges $Z$
in the range $1 \le Z \le 10$.
Note that the stability exponents are about 5 times larger in
the $eZe$ space than those in the WR; thus, trajectories approaching the DEP
will do so in general along the Wannier ridge space.

\subsection{Scattering signal for $E = 0$} \label{sec:E=0signal}
The phase space dynamics for $E = 0$ is relatively simple; the 
condition $\dot{p}_R \ge 0$ ensures in
particular that the DEP and TCP are the only fixed points 
and there are no periodic orbits and thus no chaos.
We will discuss in this section scattering signals for the $E = 0$
space in some detail and interpret them on the basis of the phase space
structure presented above. This will be helpful when turning to
the much more complex dynamics for $E < 0$ which will be investigated
in the form of a scattering problem in section
\ref{sec:E<0}.
\begin{figure}
\includegraphics[scale=0.48]{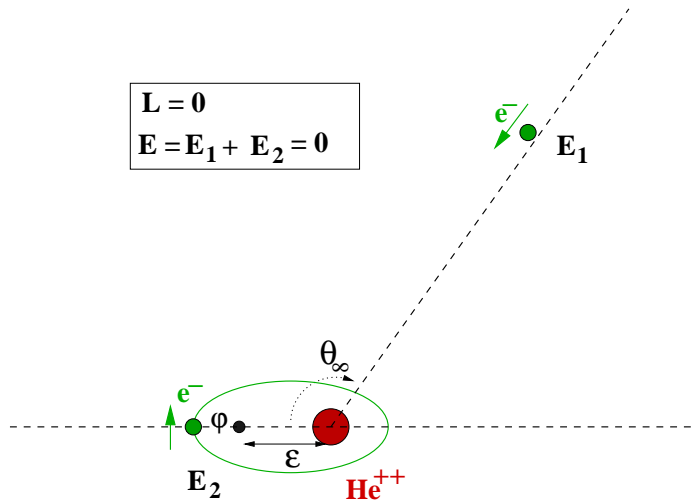}
\caption[]{\small Parametrisation of three-body Coulomb
dynamics as a scattering problem.
}
\label{fig:scat}
\end{figure}

A set of suitable parameters fully determining the initial conditions of
a scattering trajectory at energy $E=0$ and $L=0$ are shown in
Fig.\ \ref{fig:scat}; these are in particular the angle
$\theta_{\infty}$ measuring the angle between the major axis of the
Kepler-ellipse of the inner electron and the incoming direction of
the outer electron, the eccentricity $e$ of the ellipse and the
angle variable $\varphi$ of the action-angle variable pair of the
inner electron at time $t=0$. The dynamics at $E=0$ is invariant 
under changing the initial energy $E_1$ of electron 1 up to a scaling 
transformation as $E / E_1 = 0$ independent of $E_1$; we thus fix the 
$E_1 = 1$. For $e=1$ (degenerate ellipse), $\theta_\infty$ coincides with the
inter-electronic angle $\theta$ used in the hyperspherical coordinates.
The angular momentum of the incoming electron is determined
by the eccentricity $e$ and chosen such that the total angular
momentum $L=0$.  For numerical purposes, we start the incoming electron at
$r_1 = 50 Z$ and we compute the trajectory until the outgoing electron
reaches $r_i = 500Z$, $i = 1$ or $2$.

\subsubsection{The $eZe$ configuration} \label{sec:eZeE=0}
\begin{figure}
\includegraphics[scale=0.48]{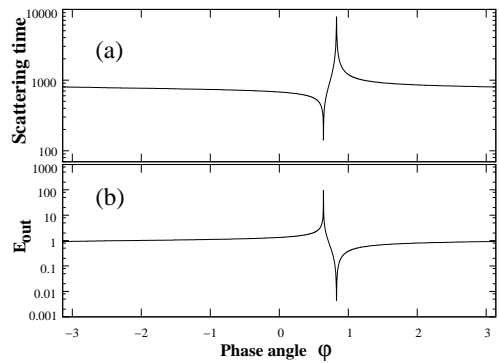}
\caption[]{\small The scattering time (unscaled) (a)
and the energy of the outgoing electron, $E_{out}$, (b)
as a function of the phase angle $\varphi$ in the collinear $eZe$
configuration ($E = 0$ and $E_1 = 1$).
Note that the scattering signals are shown on logarithmic scales.
}
\label{fig:E0eZesignal}
\end{figure}

We start with the simple case - scattering in the collinear $eZe$ space - 
for which the dynamics takes place on the boundary of the PSOS, see Fig.\ 
\ref{fig:sdep}. In Fig.\ \ref{fig:E0eZesignal}, we record the
scattering time (a) and energy of the outgoing electron (b) as a function
of the phase angle $\varphi$. The initial conditions roughly coincide with 
a cut through the $eZe$ manifold at $p_R = const \ll -P_0$.  
Note also that the scattering time is plotted here
in real time, not in the scaled time used in the McGehee transformation.

There are two exceptional orbits producing the dips and peaks at
$\varphi \approx 0.6$ and $\varphi \approx 0.8$ in the scattering time.
The dip corresponds to an initial conditions on $S_{T}^{eZe}$ and is 
thus a triple collision orbit
ending in the TCP. Orbits coming from $p_R \ll - P_0$ close
to this collision orbit will approach the TCP along the stable manifold
$S_{T}^{eZe}$ and will leave the triple collision region along the
unstable manifold $U_{T}^{eZe}$ into one of the arms leading to
single-ionisation towards $p_R \gg P_0$. The scattering time has a
minimum at that point as the escaping electron leaves with a diverging
amount of kinetic energy as one approaches the triple collision orbit,
see Fig.\ \ref{fig:E0eZesignal}(b). (Note that it takes an infinite amount of
{\em scaled} time to reach the TCP fixed point along the $S_T^{eZe}$, but as
$R \to 0$ in this limit, the unscaled momentum $p_R$ becomes singular.)

The peak in the scattering time at $\varphi \approx 0.8$ corresponds to
an orbit with initial conditions on the $S_{D}^{eZe}$ manifold
converging to the DEP fixed point and thus leading to double ionisation.
Orbits close to the $S_{D}^{eZe}$ take a large amount of scaled time
to pass the DEP which leads to large values of the hyperradius $R$.
These orbits leave the DEP region along the unstable manifold $U_{D}^{eZe}$
into one of the arms with vanishing unscaled momentum. This leads to the
dip in the energy of the outgoing electron in Fig.\ \ref{fig:E0eZesignal}(b) 
and a diverging scattering time, see Fig.\ \ref{fig:E0eZesignal}(a).
The total energy  becomes equidistributed between the two electrons
for trajectories close to the DEP; the dynamics near the TCP leads, on the
other hand, to an unequal partition of the total energy with an infinitely
fast outgoing electron and an inner electron bound infinitely deep in the
Coulomb singularity at the nucleus.

\subsubsection{Off-collinear configurations} \label{sec:90E=0}
\begin{figure}
\includegraphics[scale=0.48]{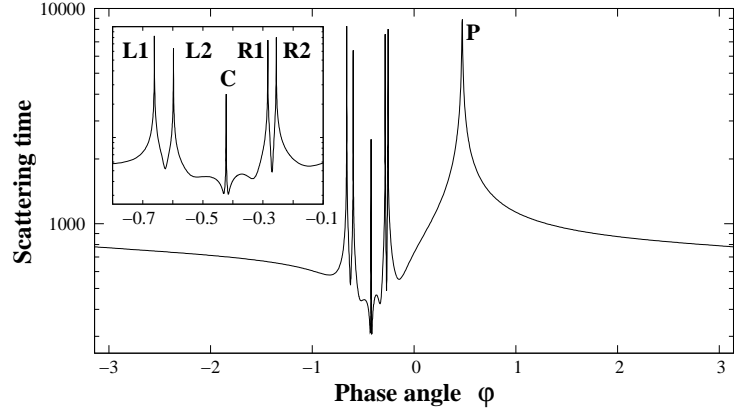}
\caption[]{\small The scattering time as a function of the phase angle
for $E = 0$ and  $e = 0.6$, $\theta_\infty = \pi / 2$, and $E_1 = - E_2 = 1$;
five distinct peaks appear in the 'dip' associated with close encounters
with the triple collision point.
}
\label{fig:90E=0Sctime}
\end{figure}

We will consider off-collinear initial conditions with $\theta_{\infty}<\pi$
next. Typical scattering signals are very similar to the one described in
the previous section for the $eZe$ configuration, see for example 
Fig.~\ref{fig:90E=0Sctime} with $e = 0.6$ and $\theta_\infty = \pi / 2$.
One finds a primary peak $P$ at $\varphi \approx 0.5$ and a dip
at $\varphi \approx -0.4$ which contains, however, a set of 5 peaks here.
To understand this signal, it is helpful to go back to the PSOS $\theta = \pi$
in Fig.\ \ref{fig:sdep}. One can identify the peak $P$ with an orbit on
the $S_D$ near the $S_D^{eZe}$ approaching the DEP 'directly' similar to
what one finds in the $eZe$ configuration.

\begin{figure}
\begin{minipage}{7cm}
\includegraphics[scale=0.38]{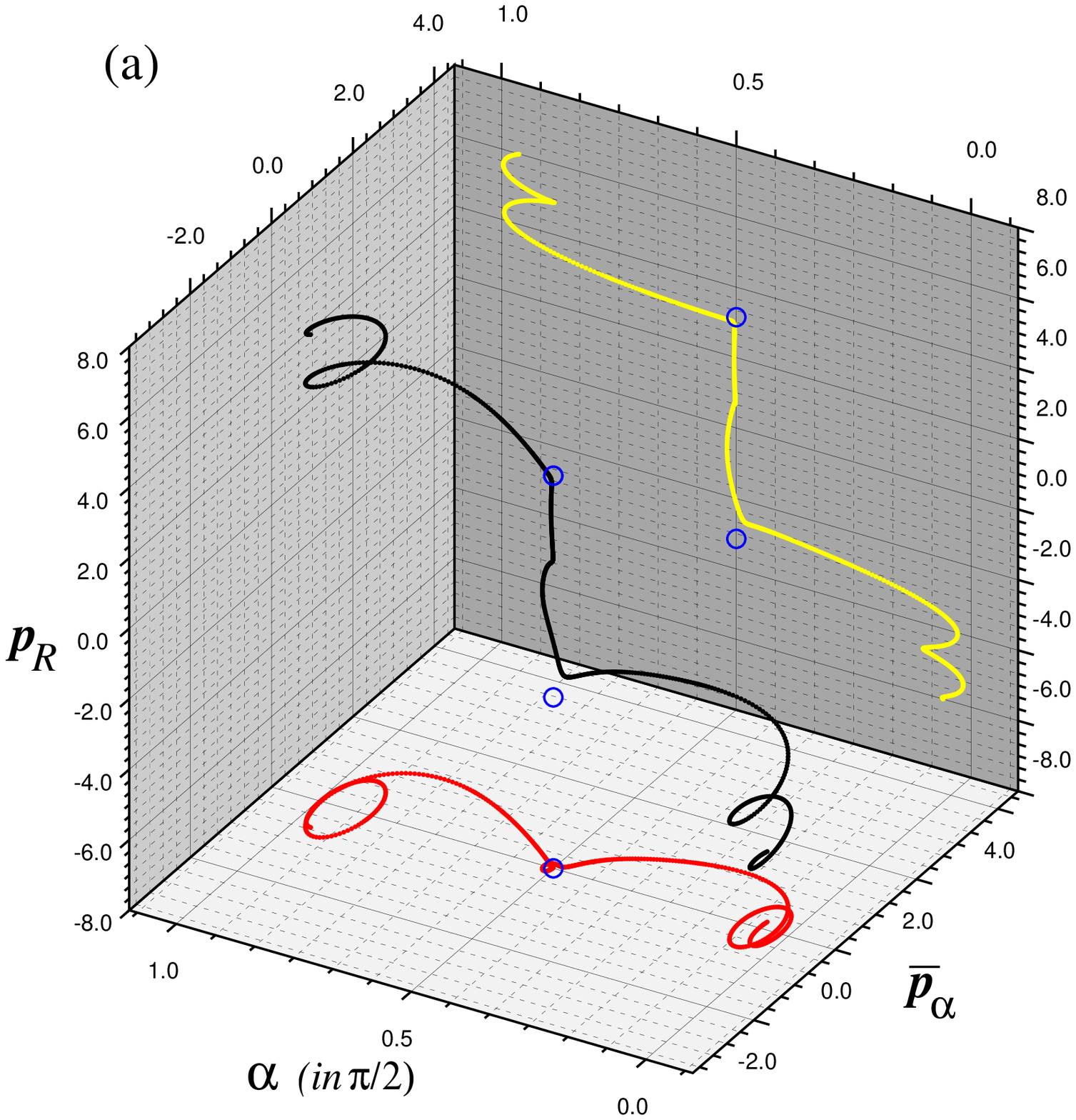}
\end{minipage}
\hfill
\begin{minipage}{7cm}
\includegraphics[scale=0.38]{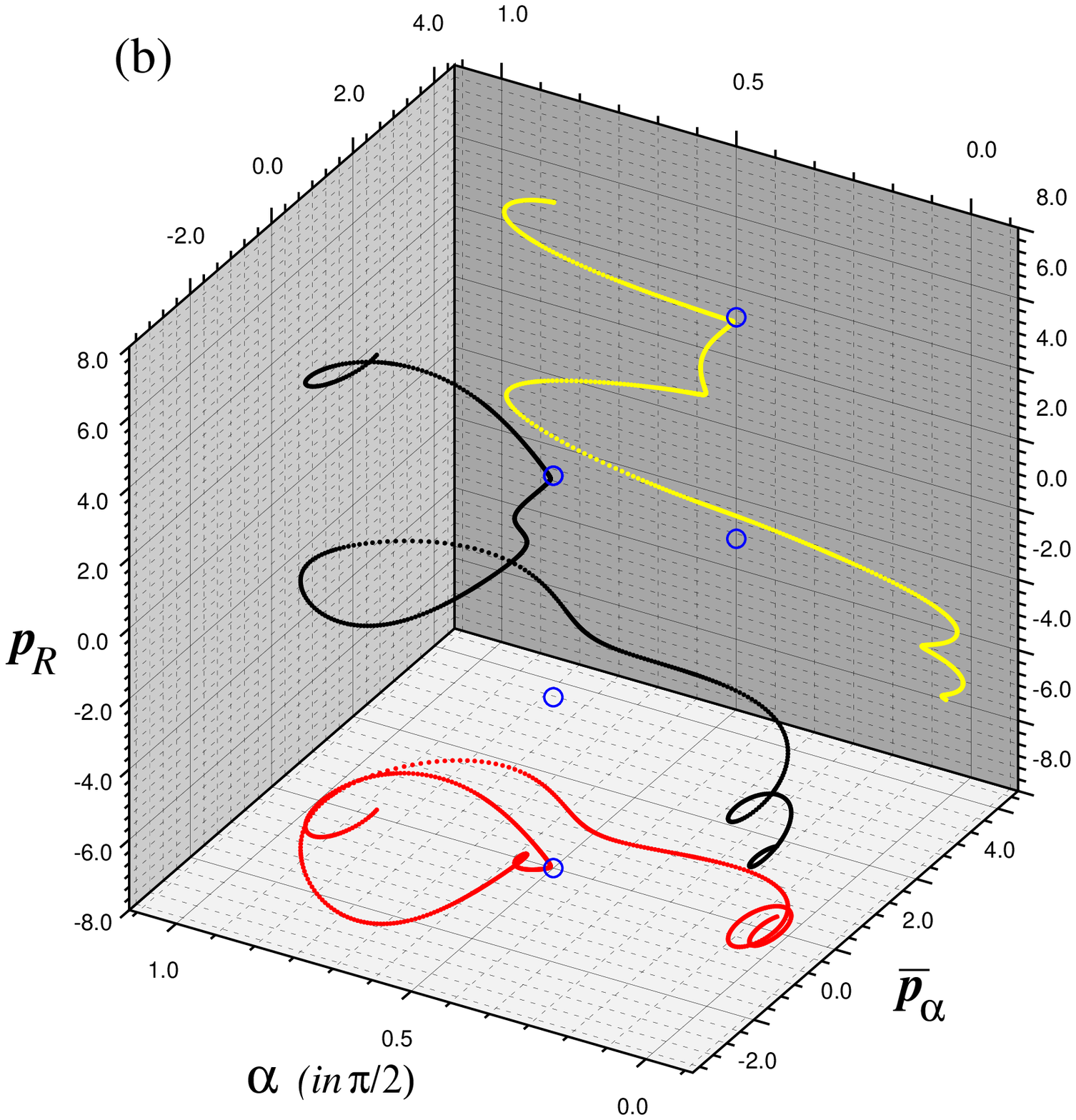}
\end{minipage}
\caption[]{\small
Scattering orbits corresponding to the C peak (a)
and the L1 peak (b), projected onto the
$\alpha - \bar{p}_{\alpha} - p_R$ space.
(Note that this is not the PSOS $\theta = \pi$, but the
full orbit where the $\theta$-dynamics has been projected
out.)  The initial conditions of the orbit are
$e = 0.6$, $\theta_\infty = \pi/2$,
$E_1 = 1$, $E_2 = -1$ with phase angles $\varphi = -0.4256$ in (a)
and $\varphi = -0.6612$ in (b) respectively.
Projections of the orbit onto the $\alpha - \bar{p}_{\alpha}$
and $\alpha -p_R$ planes are also shown. The circles represent the
positions of the DEP and TCP and their projections.
}
\label{fig:E=0Cp}
\end{figure}

New structures emerge in the dip which has in $eZe$ been associated
with a triple collision orbit on $S_{T}^{eZe}$. The TCP fixed point is,
however, no longer accessible to off-collinear initial conditions as
the stable manifold of the fixed point, $S_T$, is fully embedded in
the $eZe$ space, see Tab.\ \ref{table:table1}. Whereas near
collision orbits in $eZe$ move away from
the TCP along the unstable manifold $U_{T}^{eZe}$, another route opens up
for off-collinear orbits: escape from the TCP along the Wannier ridge which
is part of the 3 dim.\ unstable manifold $U_T$. The WR forms in fact a
heteroclinic connection between the TCP and DEP and is thus also part of the
stable manifold of the DEP, $S_D$. This and the topology of the phase
space leads to the stretching, folding, and cutting mechanism of the $S_D$
discussed in sec.\ \ref{sec:SDEP}. Orbits coming close to the TCP can thus
reach the DEP along the 5 sheets of the 3-dimensional stable manifold $S_{D}$
giving rise to the 5 peaks in the scattering signal, Fig.\ \ref{fig:90E=0Sctime}. 
The labels L1, L2, C, R1, and R2 depicted in the inset
of Fig.\ \ref{fig:90E=0Sctime} can indeed be identified with the leaves of
the $S_D$ as shown in Fig.\ \ref{fig:numbering}. The central peak, $C$,
is in particular associated with the part of the $S_{D}$
directly connected to the Wannier ridge; the outer peaks L1, L2, R2,
and R1 are related to the folded parts of the $S_{D}$ and contain orbits
which move away from the Wannier ridge after passing the TCP and before
reaching the DEP. The difference in the behaviour of the orbits in the various
leaves becomes obvious when depicting their trajectories in
$\alpha - \bar{p}_{\alpha} - p_R$ space as shown in Fig.~\ref{fig:E=0Cp}; note,
that the full orbits are shown here by projecting out the $\theta$ dynamics.
The centre-peak orbit,
Fig.~\ref{fig:E=0Cp}(a), moves indeed directly from the TCP to the DEP along
the WR which is in contrast to for example the $L_1$ orbit shown in
Fig.~\ref{fig:E=0Cp}(b).

Note that the scattering time diverges at the peaks, both for the peaks
in the dip as well as for the primary peak. The corresponding orbits
are part of the $S_D$ which is completely embedded in the
$\bar{H} = 0$ subspace. Orbits on the $S_D$ converge to the DEP and lead
thus to double ionisation.  The peaks have for $E=0$ no internal structure
which reflects the regularity of the dynamics due to the monotonic
increase of $p_R$ with time.

We have so far not discussed the $\theta_{\infty}$ - dependence on the signal.
From sec.\ \ref{sec:SDEP}, we expect that the peaks move together and converge
towards the $S_T$ as one approaches the $eZe$ boundary $\theta_{\infty}
\to \pi$, $e\to 1$. This is indeed what one observes, we will come back
to this point when discussing scaling laws in sec.\ \ref{sec:scalinglaws}.
The other limit towards the $Zee$ configuration with $\theta_{\infty} \to 0$,
$e \to 1$ is less obvious; one observes that peaks disappear in
pairs consistent with the loop - configuration of the $S_D$ as shown
in Fig.\ \ref{fig:sdep} until the scattering signal becomes flat for
small $\theta_{\infty}$. A detailed analysis of how the near $Zee$ - dynamics
is connected to the rest of the phase space will be presented in \cite{CLT05}.

\section{Dynamics for $E < 0$}
\label{sec:E<0}
We are now ready to venture into the full 5-dimensional phase space
$E<0$ with $L=0$; we will approach the problem by analysing electron
scattering signals in a similar way as in the previous section for $E=0$.
As mentioned in sec.\ \ref{sec:eq}, a smooth transition from $E < 0$ towards
$\bar{H} =0$ is achieved by taking the limit $E / E_1 \to 0$ in the initial
conditions, that is, by considering for example $E_1 \to \infty$, $E_2 \to
-\infty$ fixing the total energy at $E = -1$. In this limit, the inner electron
is bound infinitely deep in the Coulomb well and interaction between the incoming
and bound electron take place at $R \to 0$. The dynamics in $\bar{H} = 0$ is in
this sense equivalent to a dynamics at the triple collision point $R =0$.
The smooth transition implies that trajectories close to $\bar{H} =0$ will
follow the
dynamics in the $E = 0$ phase space except near the fixed points where the
flow close to the manifold $E = 0$ is perpendicular to the invariant subspace
$\bar{H}=0$ along the direction $p_R$, see (\ref{EoM}) and (\ref{stab_H}).

\begin{figure}
\includegraphics[scale=0.48]{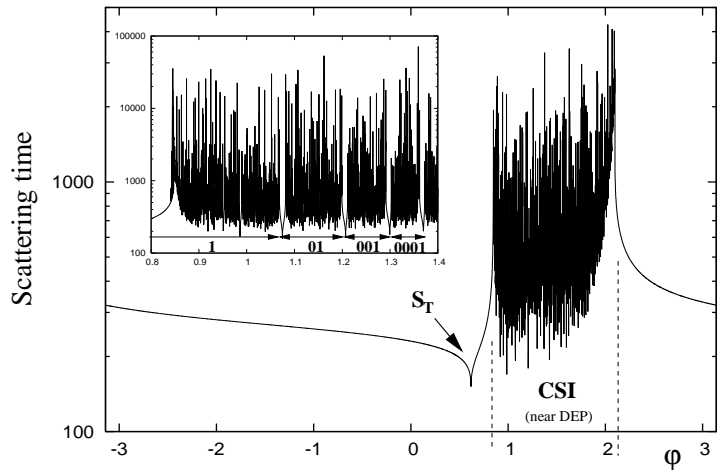}
\caption[]{\small The time-delay signal for scattering trajectories in
the $eZe$-subspace ($E_1 = 0.2$ and $E = -1$);
the intervals labelled by binary strings refer to initial conditions which
have the same symbol code after entering the chaotic scattering region
for the number of symbols given.
}
\label{fig:theta=pi}
\end{figure}

\subsection{The $eZe$ configuration} \label{sec:theta=pi}
We start again with the $eZe$ configuration which has been studied
extensively in the past \cite{Ezr91,RTW93,Ros98,Yu98,Sano04} and is
well understood by now.  Fig.~\ref{fig:theta=pi} shows the
scattering signal for $E=-1$ and $E_1 = 0.2$; compared to Fig.\ 
\ref{fig:E0eZesignal}
for $E = 0$, one finds that the peak related to the $S_D$ is replaced by
a wildly fluctuating signal typical for chaotic scattering
\cite{Gas98}. Note that the dip  related to the $S_T$ in $E=0$ is still present.

The dynamics for $E < 0$ takes place in the 3-dimensional phase space
of Fig.\ \ref{fig:fig1}a where the boundary is given by the $\bar{H} = 0$
space. The 2-dimensional stable manifold of the TCP, $S_{T}$, is
embedded in the 3-dimensional $eZe$ space spanned by
the 1-dimensional invariant manifolds $S_{T}^{\bar{H} \ne 0}$
and $S_{T}^{eZe}$; note that only the latter is in the space
$\bar{H} = 0$, see Tab.~\ref{table:table1}. The $S_T$ thus intersects the
1-dim.\ set of initial conditions for both $E = 0$ and $E\ne 0$ independent
of $E_1$. Orbits close to the $S_{T}$ manifold approach the triple collision
fixed point at $R = 0$ and follow the dynamics
along the 1-dimensional unstable manifold, $U_{T}^{eZe}$,
after passing the TCP. The dynamics here is thus similar to the one for
$E = 0$ as discussed in sec.~\ref{sec:eZeE=0};
near collision events lead to ionisation of one of the
electrons where the ionising electron escapes with a diverging amount of
kinetic energy thus giving rise to the dip in the scattering time.

The behaviour of the dynamics near the DEP is linked to the TCP-dynamics
via time reversal symmetry. The DEP is accessible only via the
stable manifold $S_{D}^{eZe}$ which is embedded in the $E=0$ space;
the DEP can thus not be reached for $E<0$ and trajectories can come
arbitrary close to the DEP only in the limit $E / E_1 \to 0$. Orbits near
the $S_{D}^{eZe}$ will, however, approach the DEP where they either
follow the flow along the unstable direction $U_{D}^{eZe}$ leading to
ionisation or follow $U_{D}^{\bar{H}\ne 0}$ (or equivalently the
Wannier orbit (WO)) into the interior of the $eZe$ space,
see Fig.\ \ref{fig:fig1}a.  In the latter case, $\dot{p}_R$
changes sign and electron trajectories fall back towards the nucleus.
The particles can now remain trapped for some time in a {\em chaotic 
scattering region} located between the TCP and DEP inside the $\bar{H} = 0$ 
manifold.  The DEP thus acts as an entrance gate into this chaotic 
scattering region. The chaotic scattering interval (CSI) in 
Fig.~\ref{fig:theta=pi} replaces the $S_D$ peak in the $E=0$ scattering 
time signal shown in Fig.\ \ref{fig:E0eZesignal};
it is directly linked to the existence of an entrance gate centred at
the DEP fixed point. By time-reversal symmetry, the TCP acts as the
exit gate for single electron ionisation.

A closer analysis of the strongly fluctuating signal in the CSI
reveals the well known binary symbolic dynamics present in the
$eZe$ configuration \cite{Yu98, Sano04, RTW93}. 
Indeed, it is now widely believed, (but still not rigorously proved), 
that the $eZe$ configuration behaves
like an ideal Smale-horseshoe, where the partition leading to a binary
symbolic dynamics is provided by the stable and unstable manifold of
the triple collision, that is, $S_T$ and $U_D$. The chaotic signal in the
CSI consists of a series of dips flanked by singularities in the delay time on
either side, see the magnified region in Fig.~\ref{fig:theta=pi}.
The dips correspond to orbits which approach the TCP along the $S_T$
after having entered the chaotic scattering region by coming close to the DEP.
Each of these triple collision orbits is embedded in an interval
of escaping trajectories, the boundaries of these intervals are given by
orbits escaping asymptotically with zero kinetic energy of the outgoing 
electron.
These orbits are thus part of the stable manifold of the asymptotic periodic
orbit where one electron stays at infinity with zero kinetic energy.
This is in contrast to the case $E = 0$ where orbits escaping with
zero kinetic energy are part of the stable manifold $S_{D}$ which leads to
double ionisation as mentioned in sec.\ \ref{sec:eZeE=0}.

\begin{figure}
\begin{minipage}{7cm}
\includegraphics[scale=0.38]{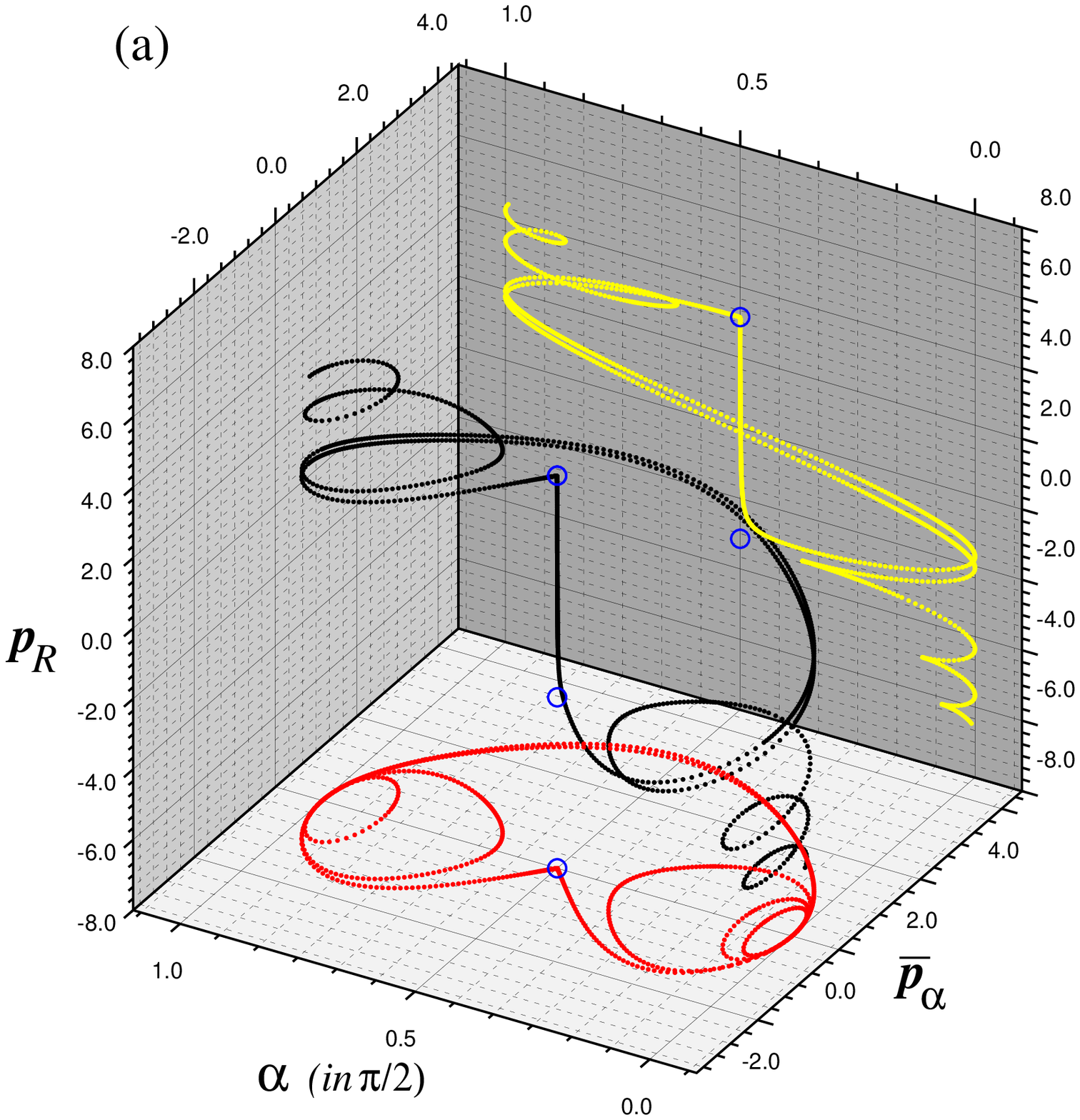}
\end{minipage}
\hfill
\begin{minipage}{7cm}
\includegraphics[scale=0.38]{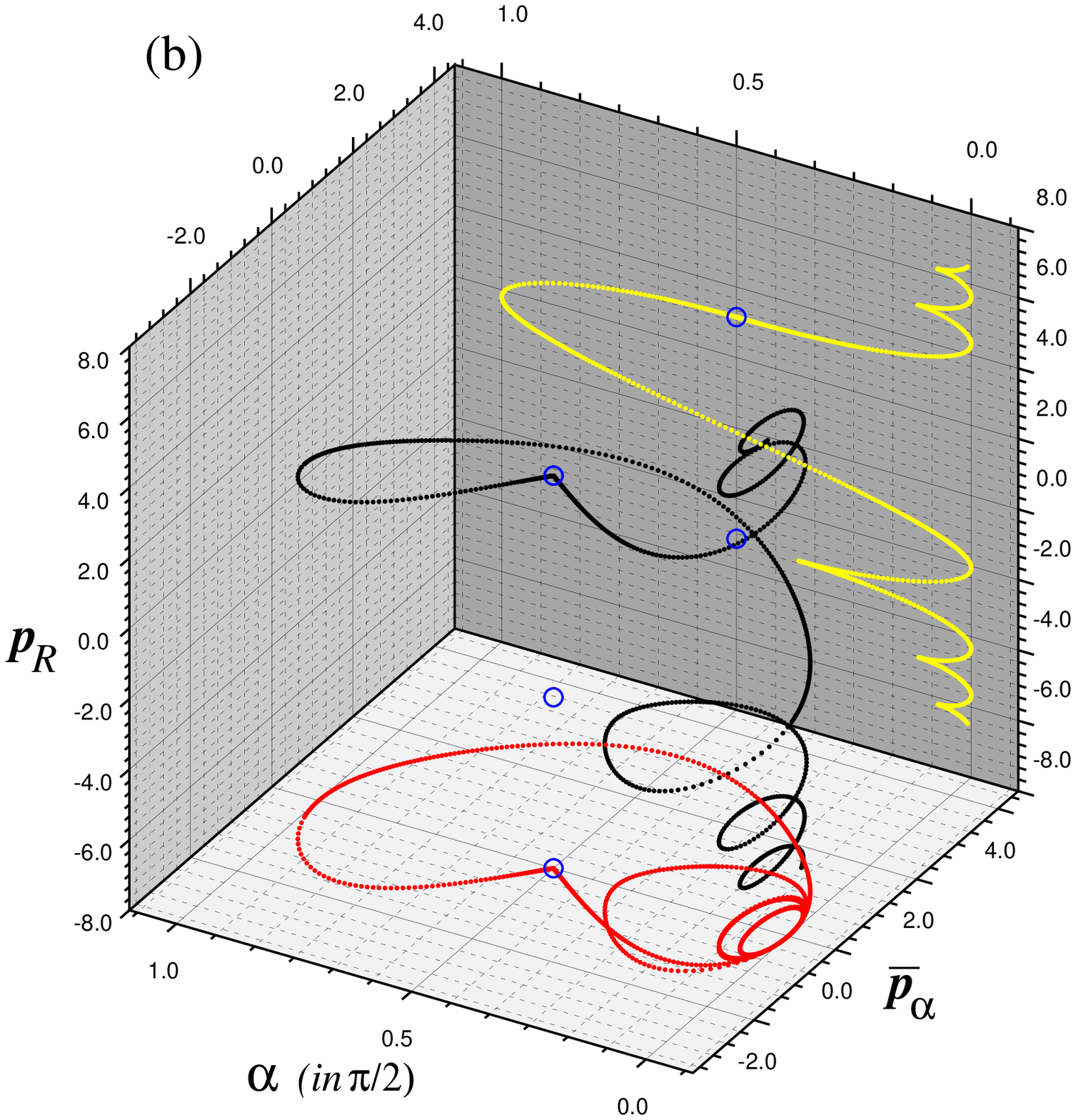}
\end{minipage}
\caption[]{\small
The shortest chaotic scattering orbit in the collinear $eZe$ space
is plotted in $\alpha - \bar{p}_{\alpha} - p_R$ coordinates for
the initial conditions $E/E_1 = -0.001$ (a); the
corresponding orbit for $E=0$ related to the peak in
Fig.~\ref{fig:E0eZesignal} is plotted for comparison in (b).
}
\label{fig:eZeorbitE=-0.001}
\end{figure}

The shortest chaotic scattering orbits correspond to the widest dip
in the CSI (see for example Fig.\ \ref{fig:theta=pi} at
$\varphi = 1.075$). The corresponding orbit for initial energy
$E_1 = 1000$ is plotted in Fig.~\ref{fig:eZeorbitE=-0.001} a  in
$\alpha-\bar{p}_{\alpha}-p_R$
coordinates. One finds indeed that the orbit approaches the DEP first
before turning towards the chaotic scattering region. In this particular
case, the orbit stays close to the WO and escapes thus immediately via
the exit gate at the TCP. An orbit close to the $S_D$ for $E=0$ is shown
in Fig.~\ref{fig:eZeorbitE=-0.001} b for comparison; this orbit
can only escape along $U_D^{eZe}$ and chaotic scattering is not possible.
Other dips in the CSI are associated with trajectories staying inside
the chaotic scattering region for longer times. The intervals between dips
can be labelled
uniquely by a finite binary code reflecting the order in which binary
collisions take place after entering and before escaping the chaotic
scattering region. We will not elaborate on the symbolic dynamics 
here, and refer the interested reader to \cite{Yu98, Sano04, RTW93}. 
Note, that the total width of the chaotic scattering interval
reduces to zero in the limit $E/E_1 \to 0$, the corresponding
scaling law is presented in sec.~\ref{sec:scalinglaws}.

\begin{figure}
\includegraphics[scale=0.48]{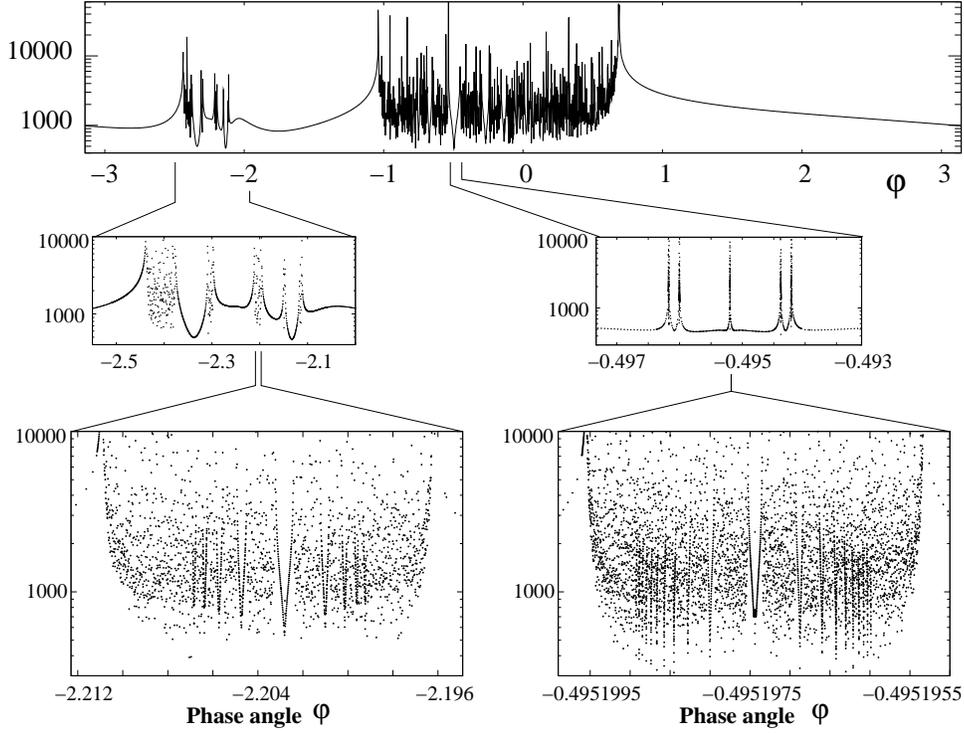}
\caption[]{\small The scattering time signal
for $\theta_\infty = \pi/2, e = 0.6$ and $E_1 = 0.2$.
}
\label{fig:90E<0Sctime}
\end{figure}
\subsection{Off-collinear configurations} \label{sec:90E<0}
From the analysis of the dynamics in the $E=0$ phase space and
the $eZe$ configuration it is now possible to understand the scattering
signals for large parts of the $E<0$ phase space by starting from the
$E/E_1 \to 0$ limit.
We note first that the stable and unstable manifolds of the triple
collision fixed points which have been so important so far are 
not contained in the off-collinear $E<0$ phase
space; indeed $S_D$ is fully embedded in $E=0$ and $ S_T$ is part of
the $eZe$ phase space, see Tab.\ \ref{table:table1}. The latter implies in
particular that triple collisions occur only in the $eZe$ configuration.
The overall dynamics is, however, clearly influenced by the invariant
manifolds of the fixed points. A typical scattering
time signal is shown in Fig.\ \ref{fig:90E<0Sctime}, here for the scattering
parameters $e = 0.6$, $\theta_\infty = \pi / 2$, $E_1 = 0.2$, and $E = -1$.
It shows a primary dip around $\varphi \approx -2.3$ containing 5 peaks
as in the off-collinear scattering data for $E=0$, see ig.~\ref{fig:90E=0Sctime},
as well as a chaotic scattering interval as in the $eZe$ case, 
see Fig.\ \ref{fig:theta=pi}.

\begin{figure}
\includegraphics[scale=0.48]{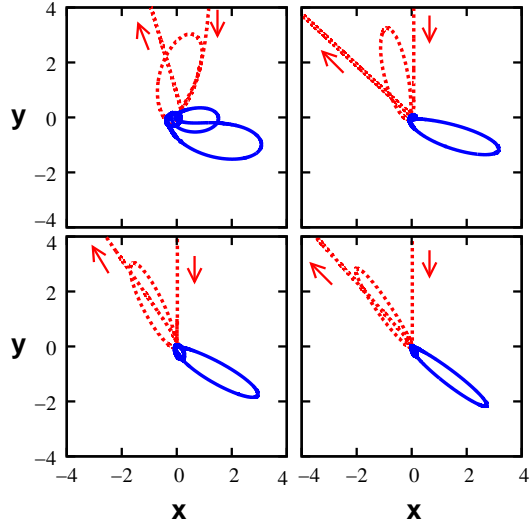}
\caption[]{\small
Short chaotic scattering orbits in the $x-y$ plane with initial
conditions in the largest dip in the primary CSI for $E_1 = 0.2$
(a), $10$ (b), $100$ (c), and $1000$ (d)
with fixed total energy $E = -1$. The full lines represent the
trajectories of the initially bound electron (with $e = 0.6$),
the dashed line correspond to the initially incoming electron
(with $\theta_\infty = \pi/2$). The nucleus is at the origin and
the direction of the semi-major axis of the initially bound electron
is aligned along the x-axis. 
}
\label{fig:variation}
\end{figure}

In analogy with the $eZe$ results, we can identify this primary CSI around
$-1.1 < \varphi < 0.7$ with the 'direct' route to the DEP close to the 
$S_D^{eZe}$. The DEP and TCP act thus again as the entrance and exit gates, 
respectively,
into or out of a chaotic scattering region. In Fig.~\ref{fig:variation},
we show a sequence of chaotic scattering orbits in configuration space
for various $E/E_1$ belonging to initial conditions in the main dip of
the CSI (such as the region around $\varphi \approx -0.495$
in Fig.\ \ref{fig:90E<0Sctime}). The trajectories pass the entrance gate
near the DEP, but leave the chaotic region immediately again by coming
close to the TCP. For small $E/E_1$, interaction between the two
electrons takes place at small values of the hyperradius $R$ and thus
close to a $R=E=0$ - dynamics; leaving the small $R$ regime into the
chaotic scattering region after passing the DEP is for $E/E_1 \to 0$
only possible along the Wannier orbit (or equivalently along 
$U_D^{\bar{H}\ne 0}$).
This can be observed in Fig.~\ref{fig:variation}d. As $E/E_1$ increases,
the trajectories move away from the WO, but retain the symmetry of the
Wannier ridge dynamics. This can be attributed to the fact, that trajectories
coming close to the DEP will do so along the Wannier ridge due to the
difference in the stability exponents along $S_D$, that is,
$\lambda_{S_{D}}^{eZe} \ll \lambda_{S_{D}}^{WR}$, see (\ref{stab_DEP}).

In contrast to the scattering signals for the $eZe$ configuration,
however, new structures appear at the centre of the dips in the CSI, 
see Fig.~\ref{fig:90E<0Sctime}. Indeed, when enlarging the intervals 
containing the dips, one finds 5 separate peaks similar to those in 
the primary 'dip' at $-2.6 < \varphi < -2.0$. In contrast to the $E=0$
case, each of these peaks is in itself a CSI on further magnification.
The origin of the 5 peaks is always the same -
close encounters with the TCP either via a direct route close
to $S_T^{eZe}$ (the primary dip) or when leaving the chaotic
scattering region (the primary CSI). The 5 peaks
can be related to the folding of the $S_D$ near the TCP
as described in section \ref{sec:SDEP}. The $S_D$ thus provides a
bridge between the TCP and DEP and trajectories can reenter the
chaotic scattering region in this way. This leads to the
secondary CSI's in each of the 5 peaks, see Fig.~\ref{fig:90E<0Sctime}.
Note that the secondary CSI's again show structures very similar to the
primary CSI and in fact similar to the CSI in the $eZe$ case.

\begin{figure}
\includegraphics[scale=0.52]{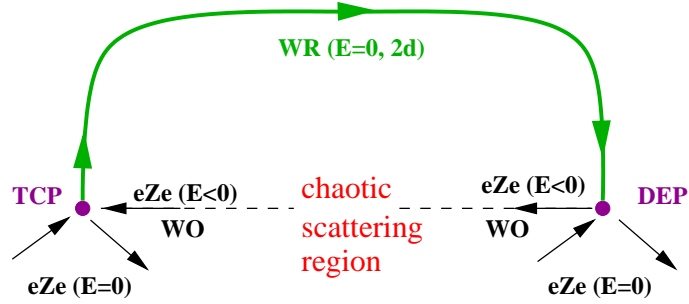}
\caption[]{\small The conveyor belt mechanism: the TCP and DEP fixed points
and their heteroclinic connections, the Wannier ridge (WR) for $E=0$ and the
Wannier orbit (WO).
}
\label{fig:TCP-DEP}
\end{figure}

The peaks in the dips suggest that it is possible to create
increasingly longer cycles of chaotic scattering events
by repeatedly moving from the DEP to the exit channel, the TCP,
and then along one of the 5 branches of the stable manifold $S_{D}$
near the TCP back to the DEP. Indeed, on further magnification
of the secondary CSI's, one finds again dips which contain 5 peaks
which on further magnification turn out to be CSI's of third order
and so on. A whole sequence of self-similar structures emerges in this 
way where dips give birth to chaotic
scattering pattern which in turn have dips containing 5 peaks etc.
The scattering data are thus a macroscopic manifestation of
the structure of the dynamics at the triple collision point.
They reflect a rather curious dynamical feature, namely a Smale-horseshoe,
whose entrance and exit points are short-circuited by two different
heteroclinic connections between the two fixed points: the Wannier ridge
(for $E=0$) leading from the TCP to the DEP and the Wannier orbit connecting
the DEP back to the TCP. This gives rise to a {\em conveyor belt} 
dynamics as it is schematically sketched in Fig.\  \ref{fig:TCP-DEP}.
 
The apparent similarities in the CSI - signals for both the collinear and
off-collinear configurations
suggests that the binary symbolic dynamics remains largely intact for a wide
range of $\theta_\infty$ values. Only the boundaries of the partition which
is formed by the $S_T$ itself in the $eZe$ case, is modified, turning into
channels from which it is possible to re-enter the chaotic scattering region.
This suggests that the 'dips' in each CSI can be labelled by a binary symbol code
related to the chaotic dynamics in the chaotic scattering region;
from here, trajectories may either escape by coming close to the TCP
or may reenter the chaotic scattering region along 5 distinct paths.
We thus expect that the dynamics can be well described in terms of $2 + 5 = 7$
symbols.\\

The analysis so far leaves many questions open. It is in particular a
big surprise that the dynamical features found in certain limits, such as the
folding of the $S_D$ in $E=0$ space or the existence of a binary symbolic dynamics
in the $eZe$ configuration, can survive in phase space regions far from these
invariant subspaces. Our numerics suggests that the conveyor belt mechanism
together with an (approximate) symbolic dynamics works in the whole range
$\pi > \theta_{\infty} > \theta_c \approx \pi/4$ and $1 > e > e_c \approx 0.6$
for energy ratios as large as $|E/E_1| = 5$. However, there must be a change in the
structure of the dynamics eventually. Results obtained in the limiting cases
$\theta_{\infty} = 0$ - the $Zee$ case \cite{RW90a} - or $e = 0$ \cite{Gu93}
certainly make this a necessity. Especially the transition from $eZe$ to $Zee$
is of importance in assigning approximate quantum numbers in (quantum)
two-electron atoms \cite{TRR00}, but remain poorly understood from a classical
mechanics point of view. The fact that the conveyor belt is so robust indicates
that there are large-scale structures in phase space at work which have not
been uncovered so far.

\subsection{Scaling laws} \label{sec:scalinglaws}
\begin{figure}
\includegraphics[scale=0.45]{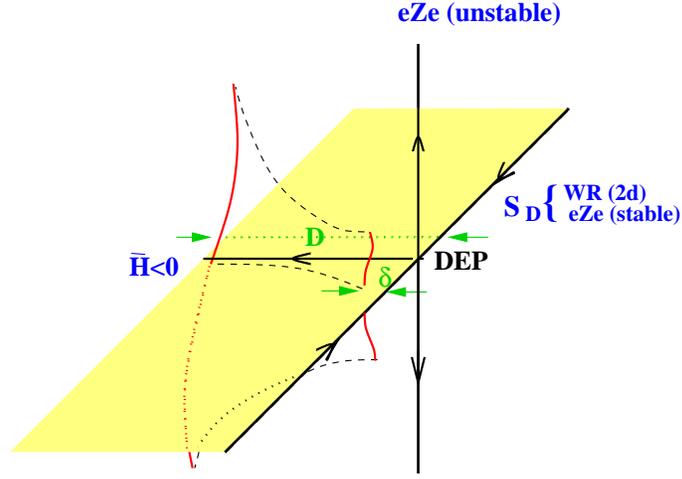}
\caption[]{\small Dynamics near the DEP}
\label{fig:DEP}
\end{figure}
Even though the scope for analytic results is limited in two-electron atom
problems, asymptotic scaling laws can be deduced from the linearised dynamics
near the fixed points. If the DEP is indeed the sole entrance gate into a 
chaotic scattering region one would in particular expect universality in the 
behaviour for
all CSI's. In the previous sections it has been argued that chaotic
scattering trajectories need to come close to the DEP before they
can flow out into the chaotic scattering region
along the unstable manifold $U_{D}^{\bar{H} \ne 0}$. In the limit
$E/E_1 \to 0$, these trajectories converge towards the $E=0$ manifold and
trajectories which will enter the chaotic scattering region along
$U_{D}^{\bar{H} \ne 0}$ need to come closer and closer to the DEP. The
phase space region which eventually enters into chaotic scattering is
limited by ejection along the other unstable manifold of the DEP,
$U_{D}^{eZe}$.

\begin{figure}
\includegraphics[scale=0.50]{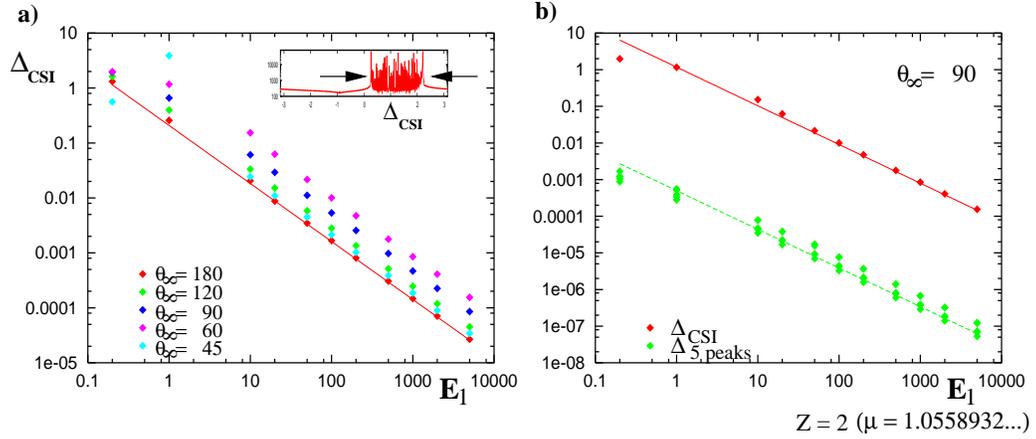}
\caption[]{\small Scaling behaviour of the width of the primary CSI
for different $\theta_{\infty}$ as well as for the 5 peaks (here for
$\theta_{\infty} = \pi/2$). The other parameters are $e=1$, $E=-1$. }
\label{fig:DEP_scalall}
\end{figure}
This implies a scaling law for the width $\Delta_{CSI}$ of the chaotic
scattering intervals for $E/E_1 \to 0$ (which should be independent of the
pre-history of these trajectory before passing the DEP entrance gate).
Let us consider the evolution of a one-dimensional set of initial condition
($-\pi < \varphi \le \pi$) for small $E/E_1$ and fixed $e$ and $\theta_\infty$.
The parts of this segment closest to the $S_D$ come close to the
DEP, see Fig.~\ref{fig:DEP}. Denote the distance from the 4-dimensional
$E=0$ manifold and thus from $S_D$ as $\delta$, that is, we have
\begin{equation} \label{delta}
\delta \propto |E/E_1| \, .
\end{equation}
Chaotic dynamics can be expected only if trajectories reach some
distance $D \approx P_0$ from the DEP along $U_{D}^{\bar{H} \ne 0}$.
The time $T_D$ for the segment to get from $\delta$ to $D$
is in linear approximation (valid for $E/E_1 \to 0$) of the order
\begin{equation} \label{T_D}
T_D \approx \frac{1}{\lambda_{U_{D}}^{\bar{H} \ne 0}}
                \log \frac{D}{\delta} \, .
\end{equation}
During that time, intervals of the size $\Delta_0$ on the segment stretch along the
$U_{D}^{eZe}$ direction according to
\begin{equation} \label{x}
\Delta(T_D) \approx \Delta_0 \exp [ \lambda_{U_{D}}^{eZe} \, T_D ]
\approx \Delta_0 (D/\delta)^\mu \, .
\end{equation}
Here,
\begin{equation} \label{mu}
\mu = \frac{\lambda_{U_{D}}^{eZe}}{\lambda_{U_{D}}^{\bar{H} \ne 0}}
= \frac{1}{4} \left(\sqrt{\frac{100 Z - 9}{4 Z - 1}} - 1\right) \,
\end{equation}
is the well known Wannier exponent controlling two-electron ionisation
processes for $E>0$ \cite{Wa53} and quantum resonance widths \cite{RW96}
near the three particle breakup threshold. 
The fraction of trajectories entering the chaotic scattering region is thus
in the limit $E/E_1 \to 0$ give  as $\Delta_{CSI} \propto \Delta_0 / \Delta(T_D)$ 
that is,
\begin{equation} \label{wannierlaw}
\Delta_{CSI} \propto (\delta/D)^{\mu} \propto \left|\frac{E}{E_1}\right|^\mu \,
\end{equation}
where the energy dependence follows from (\ref{delta}).
The scaling law is confirmed by numerical calculations and is indeed
universal, that is, it is independent of $\theta_{\infty}$, see
Fig.~\ref{fig:DEP_scalall}a, (as well as of $e$, a result not shown here),
and is the same for the primary CSI and the CSI's forming the five peaks,
see Fig.~\ref{fig:DEP_scalall}b. This clearly demonstrate that the DEP is
the sole entrance gate into the chaotic scattering region.

\begin{figure}
\includegraphics[scale=0.45]{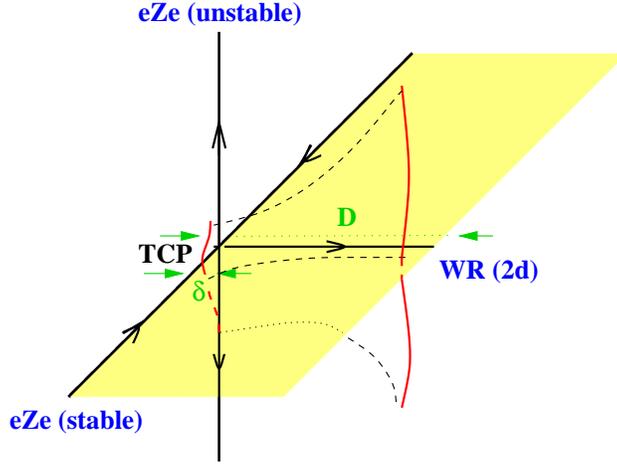}
\caption[]{\small Dynamics near the TCP}
\label{fig:TCP}
\end{figure}
\begin{figure}
\includegraphics[scale=0.65]{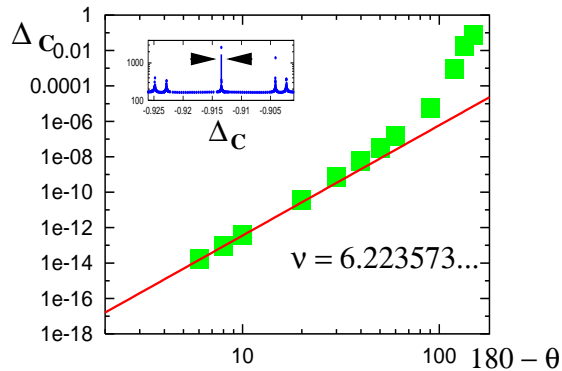}
\caption[]{\small Scaling behaviour of the centre dip as function of 
$\theta_\infty$, here for $e = 1$ and $E/E_1 = 0.1$.}
\label{fig:TCP_scal}
\end{figure}

In sections \ref{sec:90E=0}, we showed that the 5 peaks in the primary dip
are associated with 'cutting' the folded $S_D$  at the TCP which leads to
5 distinct paths from the TCP to the DEP; we argued that the centre peak
$C$ is associated with parts of the $S_D$ - manifold directly connected to
the Wannier ridge, see Figs.\ \ref{fig:sdep} and \ref{fig:numbering}.
Trajectories in the $C$-peak thus move along the Wannier ridge,
that is, along the $U_{T}^{WR}$. The phase space volume which can be 
transferred from
the TCP to the DEP along the Wannier ridge is limited by the flow
along the other unstable manifold of the TCP, $U_{T}^{eZe}$. This
implies an additional scaling law for the width of the centre peak
$\Delta_{C}$ in the limit $\theta_{\infty} \to \pi$ as well
as $E/E_1 \to 0$, (see Fig.\ \ref{fig:TCP}):
the distance of a segment of trajectories from the $eZe$ space and thus from
the TCP can be measured in terms of $\delta \propto (\pi - \theta_\infty)$.
Following a procedure similar to the derivation of (\ref{wannierlaw}),
one finds that the width of the $C$ peak interval scales as
\begin{equation} \label{tannerlaw}
\Delta_{C} \propto (\pi - \theta_\infty)^\nu \left|E/E_1\right|^\mu,
\end{equation}
with
\begin{equation} \label{nu}
\nu = \frac{\lambda_{U_{T}}^{eZe}}{{\rm Re}[ \lambda_{U_{T}}^{WR} ]}
= 1 + \sqrt{\frac{100 Z - 9}{4 Z - 1}}\, ,
\end{equation}
and $\mu$ given by (\ref{mu}).
Note, that the second part of (\ref{nu}) is valid only for
$1/4 < Z \le 9/4$; the eigenvalues $\lambda_{U_{T}}^{WR}$ become real
for $Z > 9/4$, in which case the unstable direction with the larger
eigenvalue is expected to dominate the behaviour along the WR -
coordinates.
Numerical result for the width $\Delta_{C}$ of the centre peak
in the primary dip are shown in Fig.~\ref{fig:TCP_scal} as
a function of $\pi - \theta_\infty$ for fixed $E/E_1 = -0.1$ and $e = 1$.
The agreement with the predicted scaling law demonstrates that
the centre peak is associated with the path from the TCP to the DEP 
along the Wannier ridge as described previously.

\section{Conclusions}
By using hypersherical coordinates together with McGehee scaling, it is 
possible to uncover the structure of the dynamics near the triple collision 
in detail. We first analyse the dynamics for total energy $E=0$, for which 
the set of equations of motions is reduced by one. The dynamics is here 
relatively simple compared to the $E<0$ case due to the monotonic increase 
in the momentum $p_R$ with respect to the scaled time. The DEP and TCP 
fixed point are identified as the entrance and exit
gate into and out of a chaotic scattering region within the $E<0$ space,
respectively. The two fixed points are connected along two different heteroclinic
connections, namely the  WR  for $E=0$ (going from the TCP to the DEP) and the
$WO$ for $E<0$ (connecting the DEP back to the TCP). This remarkable effect,
which has its origin in the particle exchange symmetry, together with the topology of
the phase space leads to the emergence of a 5 leaves structure of the stable manifold
of the DEP connected to the stable manifold of the TCP for $p_R < - P_0$.
This beautiful effect can be observed in scattering data for both $E=0$ and
$E<0$. In the latter case, initial conditions close to the $S_{D}$ form chaotic
scattering intervals (CSI) both for a direct route and for trajectories
near the 5 leaves of the $S_{D}$; the latter come close to the TCP before entering the
chaotic scattering region near the DEP. Scaling laws for the width of the CSI's
in the asymptotic limit $E/E_1 \to 0$ and $\theta_{\infty} \to \pi$ can be
derived in terms of the linearised dynamics near the fixed points with scaling
exponents given as ratios of stability eigenvalues.

The results describe here lay the foundations for a better understanding
of the phase space dynamics for the full 5 - dimensional phase space $E<0$. 
That there is a very robust structure becomes apparent when comparing Figs.\ 
\ref{fig:theta=pi} and \ref{fig:90E<0Sctime}. The overall signal (neglecting 
the 5 peaks) remains largely intact which suggests that the complete binary 
horseshoe spanned by the $S_T^{H\ne0}$ and the $U_D^{H\ne 0}$ in the $eZe$ 
space is continued into the full phase space. Uncovering this continuation 
process will be the key in understanding the electron-electron correlation 
effects giving rise to, for example, the existence of approximate quantum 
numbers in spectra of two-electron atoms.\\

\noindent
{\bf Acknowledgements:}\\
We would like to thank the Royal Society (GT and NNC), the Hewlett-Packard
Laboratories in Bristol (GT) and the Korea Research Foundation
(KRF-2003-015-C00119) (NNC) for financial support. Numerical support by the 
KISTI supercomputing center is thankfully acknowledged.

\begin{appendix}
\section{Non-singular equations of motion}
\label{asec:KS}
We give here the fully regularised equations of
motion in the form of a McGehee regularised version of the
3-body problem with Kustaanheimo-Stiefel (KS) regularised binary
collisions.  We follow here the treatment in ref.\cite{RTW93}, where
the regularisation of the nucleus - electron collisions
has been performed by using parabolic coordinates
for each electron which are defined by the transformations
\begin{eqnarray} \label{appendix:para}
x_1 = Q_1^2 - Q_2^2, \quad y_1 = 2 Q_1 Q_2,
\quad r_1 = R_1^2 = Q_1^2 + Q_2^2 \\ \nonumber
x_2 = Q_3^2 - Q_4^2, \quad y_2 = 2 Q_3 Q_4,
\quad r_2 = R_2^2 = Q_3^2 + Q_4^2 \\ \nonumber
p_{x_1} = \frac{Q_1 P_1 - Q_2 P_2}{2 r_1}, \quad
p_{y_1} = \frac{Q_2 P_1 + Q_1 P_2}{2 r_1}, \\ \nonumber
p_{x_2} = \frac{Q_3 P_3 - Q_4 P_4}{2 r_2}, \quad
p_{y_2} = \frac{Q_4 P_3 + Q_3 P_4}{2 r_2},
\end{eqnarray}
together with the Kustaanheimo-Stiefel time transformation \cite{KS65,AZ74}
\begin{equation} \label{appendix:KS-time}
dt = r_1 r_2 d\tau \, .
\end{equation}
Here, $(x_i, y_i)$ and $(p_{x_i}, p_{y_i})$ are
the position and momentum in Cartesian coordinates
of electron $i = 1, 2$ moving in the plane.
The notations $\bf{Q}$ and $\bf{P}$ will be used
for $(Q_1, Q_2, Q_3, Q_4)$ and $(P_1, P_2, P_3, P_4)$ respectively.
The regularised Hamiltonian $G$ can now be written as
\begin{eqnarray} \label{appendix:Hamiltonian}
   G &=&  r_1 r_2 ( H - E )  \\ \nonumber
     &=&
         \frac{1}{8} r_2 (P_1^2 + P_2^2 )  +
         \frac{1}{8} r_1 (P_3^2 + P_4^2 )   - Z r_2 - Z r_1 +
     r_1 r_2 \left( -E + \frac{1}{ r_{12} } \right) \, ,
\end{eqnarray}
where the electron-electron distance $r_{12}$ is 
\begin{equation}  \label{appendix:r12}
    r_{12} =
    \left[
          (Q_1^2 + Q_2^2)^2 + (Q_3^2 + Q_4^2)^2 -
      2 (Q_1 Q_3 + Q_2 Q_4)^2 + 2 ( Q_1 Q_4 - Q_2 Q_3)^2
     \right]^{1/2} \, .
\end{equation}
The Hamilton's equations of motion,
\begin{equation}
\frac{d \bf{Q}}{d \tau} = \frac{\partial G}{\partial \bf{P}}, \quad
\frac{d \bf{P}}{d \tau} = - \frac{\partial G}{\partial \bf{Q}} \, ,
\end{equation}
are now given as
\begin{equation} \label{appendix:EoQ}
  \frac{dQ_1 }{d \tau} = \frac{1}{4} r_2 P_1 , \quad
  \frac{dQ_2 }{d \tau} = \frac{1}{4} r_2 P_2 , \quad
  \frac{dQ_3 }{d \tau} = \frac{1}{4} r_1 P_3 , \quad
  \frac{dQ_4 }{d \tau} = \frac{1}{4} r_1 P_4 \, ,
\end{equation}
and
\begin{eqnarray} \label{appendix:EoP}
  \frac{dP_1 }{d \tau} &=&
  - \left\{
       \frac{1}{4} Q_1 ( P_3^2 + P_4^2) - 2 Z Q_1 +
       2 Q_1 r_2 \left( -E + \frac{1}{ r_{12} }\right) \right. \\ \nonumber
      &   & \left. - 2 \frac{r_1 r_2 }{r_{12}^3 }
       \left[ r_1 Q_1 + ( Q_4^2 - Q_3^2 ) Q_1 - 2 Q_2 Q_3 Q_4 \right]
    \right\} \, , \\ \nonumber
  \frac{dP_2 }{d \tau} &=&
  - \left\{
       \frac{1}{4} Q_2 ( P_3^2 + P_4^2) - 2 Z Q_2 +
       2 Q_2 r_2 \left( -E + \frac{1}{ r_{12} }\right) \right. \\ \nonumber
       &   & \left. - 2 \frac{r_1 r_2}{r_{12}^3 }
       \left[ r_1 Q_2 - ( Q_4^2 - Q_3^2 ) Q_2 - 2 Q_1 Q_3 Q_4 \right]
    \right\} \, , \\ \nonumber
  \frac{dP_3 }{d \tau} &=&
  - \left\{
       \frac{1}{4} Q_3 ( P_1^2 + P_2^2) - 2 Z Q_3 +
       2 Q_3 r_1 \left( -E + \frac{1}{ r_{12} }\right) \right. \\ \nonumber
       &   & \left. - 2 \frac{r_1 r_2}{r_{12}^3 }
       \left[ r_2 Q_3 + ( Q_2^2 - Q_1^2 ) Q_3 - 2 Q_1 Q_2 Q_4 \right]
    \right\} \, , \\ \nonumber
  \frac{dP_4 }{d \tau} &=&
  - \left\{
       \frac{1}{4} Q_4 ( P_1^2 + P_2^2) - 2 Z Q_4 +
       2 Q_4 r_1 \left( -E + \frac{1}{ r_{12}}\right) \right. \\ \nonumber
       &   & \left. - 2 \frac{r_1 r_2}{r_{12}^3 }
       \left[ r_2 Q_4 - ( Q_2^2 - Q_1^2 ) Q_4 - 2 Q_1 Q_2 Q_3 \right]
    \right\} \, .
\end{eqnarray}
Singular behaviour may occur at the triple collision and thus in terms 
containing $1/r_{12}$ in (\ref{appendix:EoP}); it turns out, however, that
the KS - time transformation (\ref{appendix:KS-time}) also lifts the
triple collision singularity from the equations of motion: $r_{12} = 0$
can indeed only occur if $r_1 = r_2 = 0$ due to the $e - e$ repulsion.
Terms containing $1/r_{12}$ in (\ref{appendix:EoP})
indeed vanish proportional to $\sqrt{R}$ when $r_{12} \to 0$ where the
hyperradius $R$ in the new coordinates takes on the form
\begin{equation}
R = \sqrt{(Q_1^2 + Q_2^2)^2 + (Q_3^2 + Q_4^2)^2}.
\end{equation}

It is, however, still advantageous to employ McGehee - scaling as introduced in 
sec.~\ref{sec:eq} in addition to a KS - transformation. By defining
\begin{equation} \label{appendix:scaledQ}
\bar{\bf{Q}} = \frac{\bf{Q}}{\sqrt{R}} \, ,
\end{equation}
one arrives at a set of coordinates where the $\bar{\bf{Q}}$ and
$\bar{r}_{12}$ can take on non-zero values at the triple collision.
(One can actually show that $\bar{r}_{12} > 0$ everywhere for $E \le 0$,
for example $\bar{r}_{12} > 0.156\ldots$ for $Z=2$.) This means
in particular that expressions containing $r_{12}$ in
(\ref{appendix:EoP}) remain in general finite in the
McGehee - scaled coordinates even at the triple collision.
For numerical calculations, it is thus more convenient to use
the scaled coordinates which are less sensitive to numerical
errors due to small denominators.

After introducing the additional time transformation
\begin{equation} \label{tildetau}
d \bar{\tau} = \sqrt{R} \, d \tau \, ,
\end{equation}
(which leads to a speed-up near the triple collision compared to
using KS - time transformation only),
one obtains the equations of motion for the scaled coordinates
$\bar{\bf{Q}}$ as
\begin{eqnarray}  \label{appendix:scaledEoQ}
  \frac{d\bar{Q}_1 }{d \bar{\tau}} &=&
     \frac{1}{4} \bar{r}_2 \bar{P}_1
    -\frac{1}{2} \bar{Q}_1 \bar{r}_1 \bar{r}_2 \bar{p}_R \,
     \\ \nonumber
  \frac{d\bar{Q}_2 }{d \bar{\tau}} &=&
     \frac{1}{4} \bar{r}_2 \bar{P}_2
    -\frac{1}{2} \bar{Q}_2 \bar{r}_1 \bar{r}_2 \bar{p}_R \,
     \\ \nonumber
  \frac{d\bar{Q}_3 }{d \bar{\tau}} &=&
     \frac{1}{4} \bar{r}_1 \bar{P}_3
    -\frac{1}{2} \bar{Q}_3 \bar{r}_1 \bar{r}_2 \bar{p}_R \,
     \\ \nonumber
  \frac{d\bar{Q}_4 }{d \bar{\tau}} &=&
     \frac{1}{4} \bar{r}_1 \bar{P}_4
    -\frac{1}{2} \bar{Q}_4 \bar{r}_1 \bar{r}_2 \bar{p}_R \, ,
\end{eqnarray}
with
\begin{equation} \label{appendix:scaledP}
\bar{\bf{P}} = \bf{P} \, ,
\end{equation}
and
$\bar{p}_R$ is the scaled momentum of the hyperradius as in
(\ref{scal_R}); it can be expressed in terms of the scaled
parabolic coordinates as
\begin{equation}  \label{appendix:p_R}
\bar{p}_R = \frac{1}{2}
             (\bar{Q}_1 \bar{P}_1 +
             \bar{Q}_2 \bar{P}_2 +
             \bar{Q}_3 \bar{P}_3 +
             \bar{Q}_4 \bar{P}_4 ) \, .
\end{equation}
The equations of motion for $\bar{\bf{P}}$
are the same as in (\ref{appendix:EoP})
where the variables $\bf{Q}$, $\bf{P}$, $\tau$  and the energy $E$
are replaced by the scaled variables;
the scaled energy $\bar{E}$ is as in (\ref{scal_R}) given by
\begin{equation} \label{appendix:scaledE}
\tilde{E} = R E  \,\quad \mbox{with} \quad
\frac{d \bar{E}}{d \bar{\tau}} = \bar{r}_1 \bar{r}_2
\bar{p}_R \bar{E} \, .
\end{equation}
The full set of equations of motion
(\ref{appendix:EoP}), (\ref{appendix:scaledEoQ}), and (\ref{appendix:scaledE})
are free of singularities and are numerically stable both in the vicinity of
binary and triple collisions.

\section{Poincare surface of section}
\label{appendix:sec:PSOS}
A 'good' global Poincare surface of section (PSOS) should fulfil two
basic ingredients, namely
\begin{itemize}
\item[i)] almost all trajectories cross the PSOS;
\item[ii)] the vector-field is transversal to the PSOS (except on
lower dimensional invariant manifolds).
\end{itemize}
The latter condition is readily fulfilled for the PSOS $\theta = \pi$
as
\[
\dot{\theta} = \frac{p_{\theta}}{\sin^2\alpha \cos^2\alpha} \ne 0
\]
for all points on the PSOS except those in the invariant $eZe$ space
with $\theta = \pi, p_\theta = 0$.
 
\begin{figure}
\includegraphics[scale=0.40]{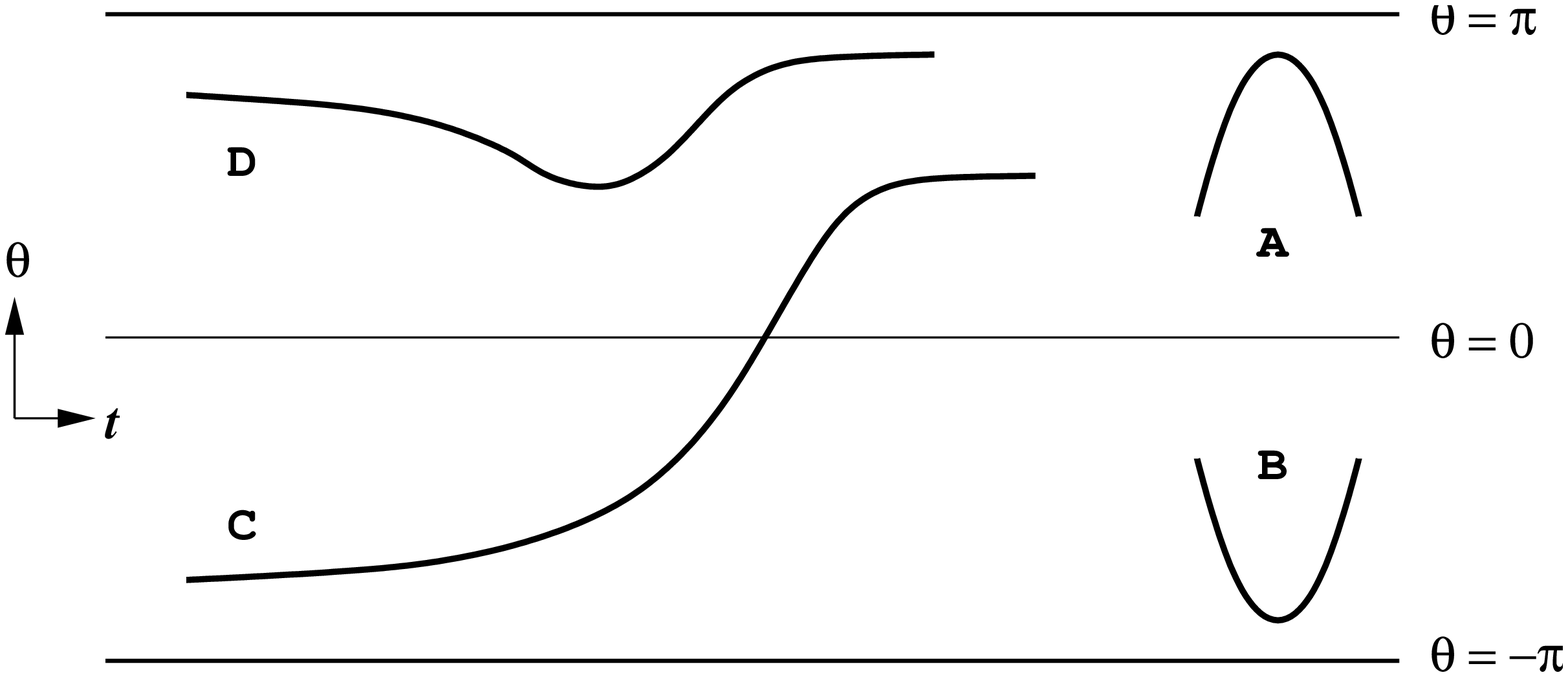}
\caption[]{\small Impossible trajectories.
}
\label{app:fig1}
\end{figure}

Next, we show that a generic orbit with total angular momentum $L = 0$
intersects the hyper-surface $\theta = \pi$ in all three energy regimes
$E =0, \pm 1$ at least once. Let us assume that there are trajectories
which never intersect $\theta = \pi$ for all times. A possible way
for this to happen is, that trajectories oscillate in the range
$\theta \in [-\pi,\pi]$ without crossing the PSOS. This means, there
must be turning points of the form $\bf{A}$ and $\bf{B}$ in Fig.~\ref{app:fig1},
where $p_\theta= 0$ with $-\pi < \theta <\pi$. However, employing
(\ref{EoM}), we have at such a point
\begin{equation} \label{appendix:ptheta}
\dot{p}_\theta =
= \frac{\sin  2 \alpha  \sin \theta}{2 [1 - \sin 2 \alpha  \cos \theta]^{3/2}}
             \left\{
              \begin{array}{ccl}
                 < 0  & {\rm for}& -\pi  < \theta  < 0 \; (\bf{B})   \\
                 > 0  & {\rm for}& 0 < \theta  <  \pi  \; (\bf{A})
              \end{array}
             \right. \, ,
\end{equation}
whereas we would need $\dot{p}_\theta < 0$ in scenario $\bf A$ and
$\dot{p}_\theta > 0$ in $\bf B$. These cases can thus be excluded.

The other possibility is that there exist trajectories which never cross the
PSOS by converging to a fixed value in $\theta$ with $\theta \ne 0$ or $\pi$
and thus $p_\theta \to 0$ for $t \to \pm \infty$ as indicated by the cases $\bf{C}$
and $\bf{D}$ in Fig.~\ref{app:fig1}. (Orbits converging towards $\theta =0$ or $\pi$
must lie at homoclinic or heteroclinic intersections of the stable and unstable
manifolds of the invariant $Zee$ or $eZe$ subspaces and are thus of measure
zero in the full phase space.) If convergence in $\theta$ occurs for
$t \to \pm \infty$, this implies $\dot{p}_{\theta} \to 0$ and thus
$\alpha \to 0$ or $\pi/2$ in these limits, see
(\ref{appendix:ptheta}). Furthermore, from eqn.\ (\ref{angular}) we have
$p_{\theta} =p_{\theta_1} = - p_{\theta_2}\to 0$, that is, both electrons
have angular momentum zero asymptotically. This is possible only if
$\theta = 0$ or $\pi$ or if one of the two electrons escapes to infinity.
The final state must thus be an incoming and outgoing scattering trajectory
of the type shown in Fig.\ \ref{fig:scat} with $\epsilon = 1$. However,
for finite $\alpha$, the electron-electron interaction will push the inner
electron onto an elliptic motion around the nucleus and $\theta$ will thus
cross $\theta = \pi$. This gives the contradiction and there are no trajectories
of the form $\bf{C}$ and $\bf{D}$ as depicted in Fig.~\ref{app:fig1}.
Consequently the hyper-surface $\theta = \pi$ is a suitable Poincare surface of
section for all energies.

\end{appendix}

\end{document}